\documentclass{iopart}
\usepackage{iopams}
\usepackage{graphicx}
\def\beq{\begin{equation}}
\def\eeq{\end{equation}}
\def\beqa{\begin{eqnarray}}
\def\eeqa{\end{eqnarray}}

\def\bN{\mathbf N}
\def\N{ {\cal N} }

\def\L{\mathrm{L}}
\def\R{\mathrm{R}}

\def\r{\mathbf{r}}

\def\j{\mathbf{j}}
\def\J{\hat{J}}
\def\F{\hat{F}}
\def\K{\hat{K}}
\def\v{\boldsymbol v}
\def\s{\boldsymbol s}
\def\u{\boldsymbol u}

\begin{document}

\title{Semiclassics for chaotic systems with tunnel barriers}

\author{Jack Kuipers}

\address{Institut f\"ur Theoretische Physik, Universit\"at Regensburg, D-93040 Regensburg, Germany}
\ead{Jack.Kuipers@physik.uni-regensburg.de}
\begin{abstract}
{The addition of tunnel barriers to open chaotic systems, as well as representing more general physical systems, leads to much richer semiclassical dynamics.  In particular, we present here a complete semiclassical treatment for these systems, in the regime where Ehrenfest time effects are negligible and for times shorter than the Heisenberg time.  To start we explore the trajectory structures which contribute to the survival probability, and find results that are also in agreement with random matrix theory.  Then we progress to the treatment of the probability current density and are able to show, using recursion relation arguments, that the continuity equation connecting the current density to the survival probability is satisfied to all orders in the semiclassical approximation.  Following on, we also consider a correlation function of the scattering matrix, for which we have to treat a new set of possible trajectory diagrams.  By simplifying the contributions of these diagrams, we show that the results obtained here are consistent with known properties of the scattering matrix.  The correlation function can be trivially connected to the ac and dc conductances, quantities of particular interest for which finally we present a semiclassical expansion.}
\end{abstract}


\section{Introduction} \label{intro}

Quantum systems that are chaotic in the classical limit exhibit universal behaviour that can be well modelled by random matrix theory (RMT) which involves treating the Hamiltonian as a matrix with random elements \cite{mehta04,bgs84,haake00}.  Another approach is to use semiclassical methods to obtain approximations in terms of classical trajectories, which are valid in the semiclassical limit $\hbar\to0$.  These methods, as well as providing an explanation of the observed universal behaviour in terms of classical correlations, can also be used to explore system-specific quantum properties \cite{stockmann99,bb03}.

The semiclassical trajectory based techniques which we use here were first developed to treat the spectral form factor $K(\tau)$, which for quantum chaotic systems has a universal form depending only on the symmetries of the system and which can be obtained from RMT.  The form factor can be written semiclassically using Gutzwiller's trace formula \cite{gutzwiller71,gutzwiller90} as a double sum over periodic orbits.  By pairing periodic orbits with themselves (or their time reverse), known as the `diagonal' approximation, it was shown \cite{berry85}, using the sum rule of \cite{ha84}, that this recreated the leading order term in the small $\tau$ expansion of the form factor.  

\begin{figure}
\centering
\includegraphics[width=8cm]{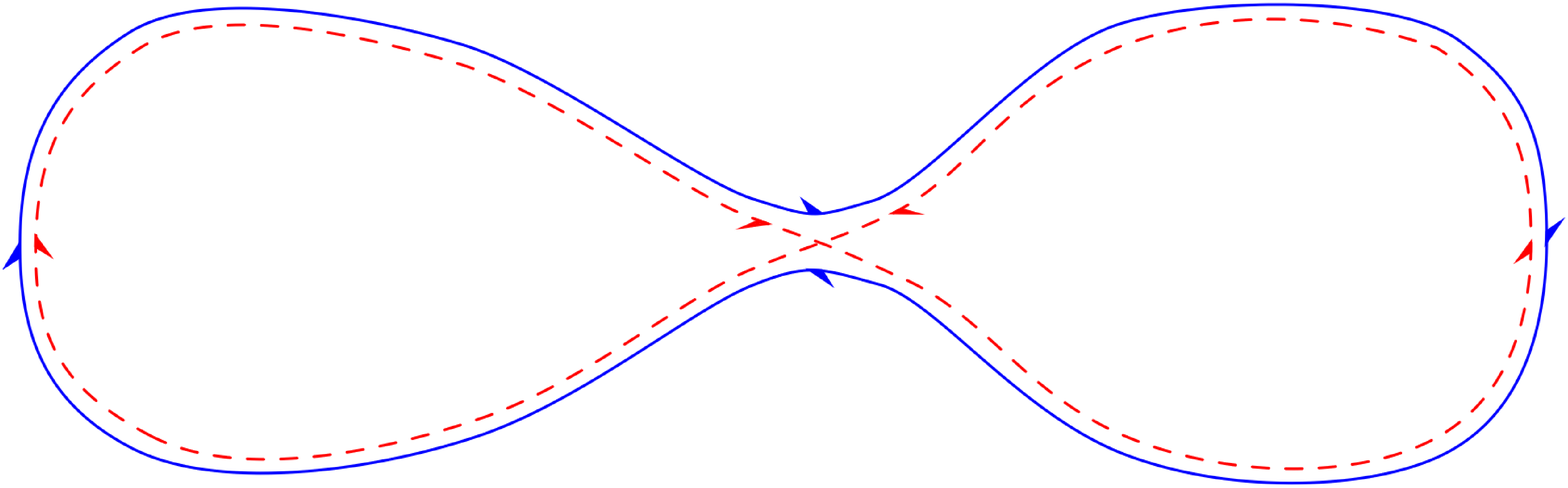}
\caption{The type of periodic orbit pair that gives the first off-diagonal contribution to the spectral form factor for systems with time reversal symmetry.}
\label{goeencounterpic}
\end{figure}

Contributions beyond the diagonal approximation come from pairs of correlated periodic orbits whose action difference is small on the scale of $\hbar$.  The first such pair was found in \cite{sr01} for a system with uniformly hyperbolic dynamics, and is depicted schematically in figure~\ref{goeencounterpic}.  It consists of an orbit with a small angle self crossing and a partner that follows almost the same trajectory, but which avoids crossing and completes the trajectory back to the crossing in the opposite direction.  Such pairs can therefore only exist in systems with time reversal symmetry, and were shown to give the first off-diagonal correction to the form factor, which agrees with the second order term of the orthogonal random matrix result \cite{sr01,sieber02}.

\begin{figure}
\centering
$\begin{array}{ccc}
a)
\includegraphics[width=4cm]{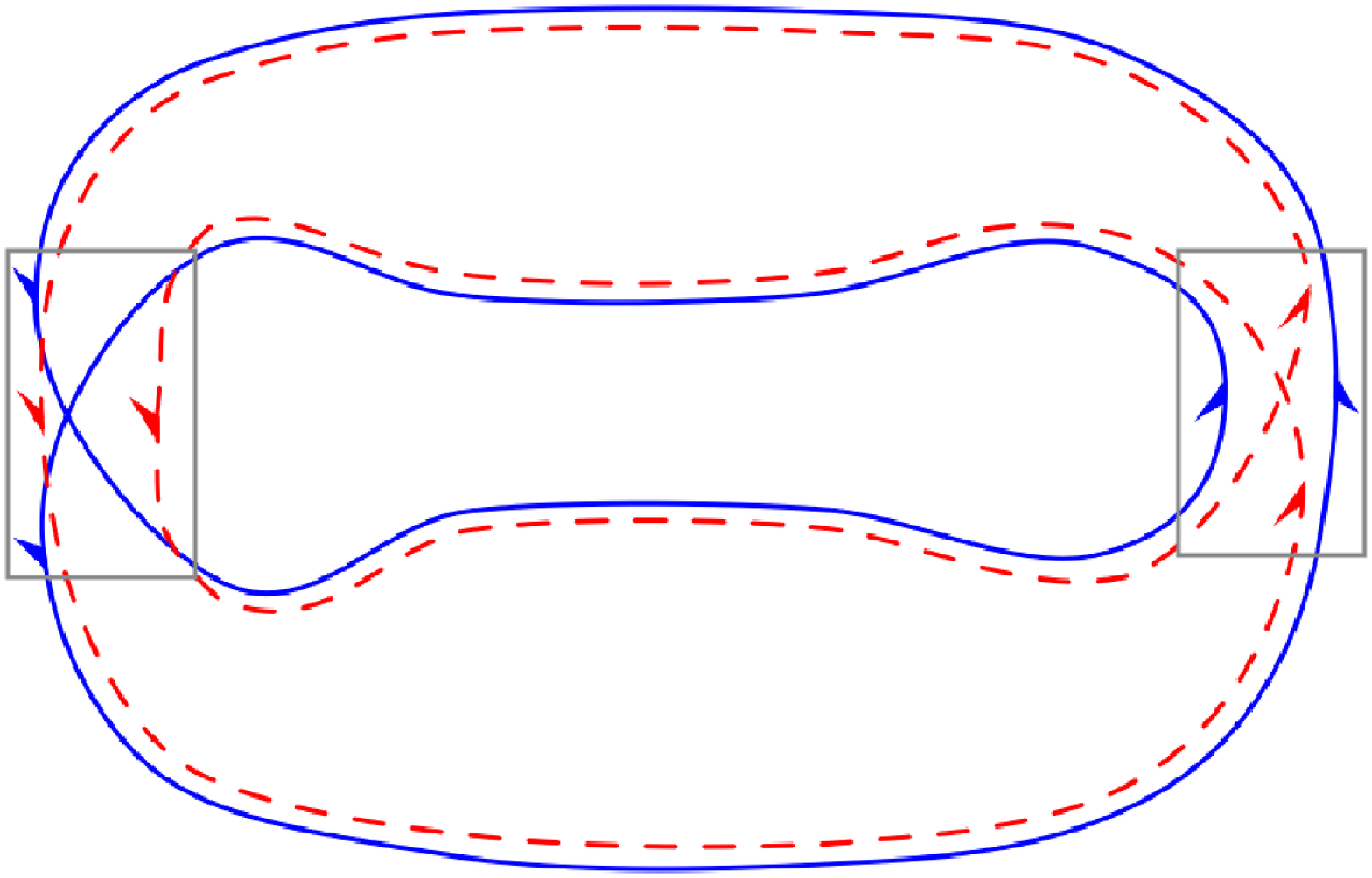} & \;\;
& b) \includegraphics[width=4cm]{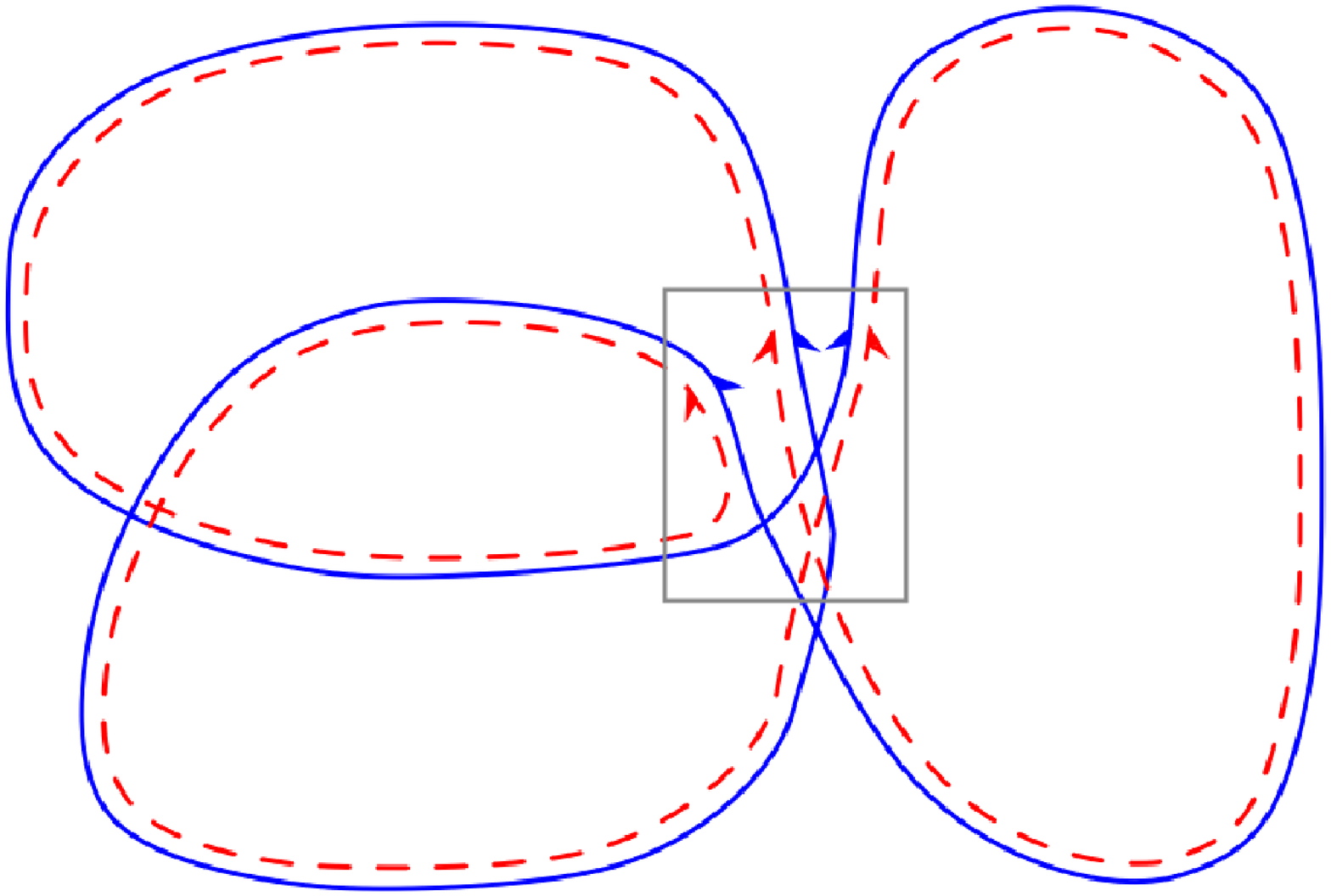}
\end{array}$
\caption{The types of periodic orbit pairs that give the first off-diagonal contributions to the form factor for systems without time reversal symmetry.  The encounter regions are indicated by the rectangles.}
\label{gueencounterpic}
\end{figure}

Not long after these ideas were reformulated in terms of phase space coordinates instead of crossing angles \cite{spehner03,tr03,tureketal05}, the orbit pairs responsible for the next order correction were identified \cite{heusleretal04, heusler03}, and their contribution shown to agree with the next term in the RMT result.  Of these orbit pairs, those that are possible for systems without time reversal symmetry are depicted in figure~\ref{gueencounterpic}.

These ideas and calculations were further extended in \cite{mulleretal04,mulleretal05} to cover orbits with an arbitrary number of encounters each involving an arbitrary number of stretches.  A self-encounter that involves $l$ stretches of the trajectory is called an `$l$-encounter', and the encounter stretches are separated by long trajectory stretches called `links'.  By using the hyperbolicity and long time ergodicity of the chaotic dynamics, as well as considering the number of different possible configurations of orbit pairs, they were able to generate all terms of the small $\tau$ RMT expansion for the unitary, orthogonal and symplectic symmetry classes \cite{mulleretal04,mulleretal05, muller05b}.  Since then, these methods have been successfully applied to show agreement with RMT, for $\tau<1$, for the transition between the unitary and orthogonal symmetry classes, parametric correlations, open systems and combinations of all of these \cite{sn06,nagaoetal07,ks07a,ks07b}.

Recently though, exciting progress in the regime $\tau>1$ has come from these types of methods by considering a generating function of the correlation function using spectral determinants \cite{heusleretal07,mulleretal09}.  This, along with the use of resummation \cite{km07} which re-expresses the sum over long periodic orbits in terms of shorter ones, allowed the correlation function to be expressed in terms of a sum over four sets of periodic orbits (or pseudo or composite orbits) and then the full form factor for all $\tau$ to be recreated.

On a different front, the application of these methods to quantum transport follows a similar history as for periodic orbit correlations.  One quantity of particular interest is the conductance \cite{fl81}, which is given semiclassically by trajectories which start in one lead and travel to another \cite{miller75,bs89,richter00}.  The diagonal terms were evaluated first \cite{bjs93a,bjs93b}, before the first off-diagonal contributions were identified \cite{rs02}.  These are related to the periodic orbit pairs in figure~\ref{goeencounterpic} and can be formed by cutting the link on the left and moving the cut ends to the leads.  Identifying the connection between the possible periodic orbit structures and the open trajectory structures, and building on their work on spectral statistics, the semiclassical expansion for the conductance was then calculated to all orders \cite{heusleretal06}.  The same treatment was successfully applied to the shot noise \cite{braunetal06}, conductance fluctuations and other correlation functions \cite{mulleretal07}, giving results that agreed with RMT (where RMT predictions existed) and a complete semiclassical treatment of quantum transport in that regime.  These methods have then been applied to include spin interactions \cite{bw07}, to treat the time delay \cite{ks08}, and to provide the leading order contributions to higher order correlation functions \cite{bhn08}.

However, the power of semiclassical methods is not just restricted to recreating RMT results, since they can also be applied to regimes where RMT no longer holds.  One example is the periodic orbit encounters treated in \cite{ks08, kuipers08b} which semiclassically recreate the oscillatory terms in the time delay, while a large area of interest is in Ehrenfest time effects.  For the conductance, the Ehrenfest time dependence of weak localisation \cite{al96,adagideli03} and coherent backscattering \cite{jw06,rb06} has been treated semiclassically.  Beyond this, the Ehrenfest time dependence has also been found for the leading order of the shot noise \cite{wj06}, conductance fluctuations \cite{br06} and a third order correlation function \cite{br06b}.  This work has shown that we need to treat additional trajectory diagrams that only play a role when the Ehrenfest time is important.  These include the coherent backscattering contribution, which has an Ehrenfest time dependence and can no longer be treated as part of the diagonal approximation, as well as the periodic orbit encounters which appear for the conductance fluctuations \cite{br06,brouwer07}.

The semiclassical treatment for the conductance and related quantities involves trajectories that start and end in the lead, so (apart from coherent backscattering) there can be no encounters at the end of the trajectories as the trajectory must escape and can no longer return to a nearby point.  However, when tunnel barriers were included in \cite{whitney07}, because the semiclassical trajectories can be reflected when they try to escape, encounters can now occur at the leads and additional diagrams exist.  Indeed, the `failed' coherent backscattering and other extra diagrams are necessary to preserve the unitarity of the evolution \cite{whitney07}.  That work was concerned with the leading order corrections when Ehrenfest time effects are important, but leaving behind this regime similar types of diagrams were also developed to treat the survival probability \cite{waltneretal08}.  As the semiclassical approximation for the survival probability involves trajectories that start and end inside the system, encounters can occur near the start or the end of the trajectory, leading to the `one-leg-loop' diagrams of \cite{waltneretal08}.  These were generalised in \cite{gutierrezetal09} where the relationship between these new diagrams and closed periodic orbit structures was also explored.  These connections were further explored in the context of the semiclassical continuity equation in \cite{kuipersetal08} where combinatorial arguments were used to show that the continuity equation is satisfied in the semiclassical approximation.  This is the basis of the current work, and the goal of this article is to combine semiclassical trajectory based expansions \cite{sr01,mulleretal04,mulleretal05,rs02,heusleretal06,mulleretal07}, with the treatment of tunnel barriers \cite{whitney07} and including the extra diagrams and their generalisations \cite{waltneretal08, gutierrezetal09}, to provide a complete semiclassical treatment for chaotic systems with tunnel barriers, at least in the regime where Ehrenfest time effects are negligible and for times shorter than the Heisenberg time.

To do this we first examine the changes \cite{whitney07} that adding tunnel barriers brings to the system in section~\ref{tunnels}.  We then re-examine the semiclassical continuity equation \cite{kuipersetal08}, which allows us to explore all the different types of diagrams that contribute, and concentrate first on the survival probability in section~\ref{survprob}.  To simplify the semiclassical treatment we shift to the Fourier domain in section~\ref{fourier}, and then use recursion relation arguments that allow us to combine sums of semiclassical contributions in an elegant way.  The probability current density, which we treat in section~\ref{current}, is related to the survival probability via the continuity equation, and after again simplifying the contributions, we show that the continuity equation is satisfied to all orders with the presence of tunnel barriers.  In section~\ref{transport}, we consider a correlation function of the scattering matrix, for which we have to treat a new set of possible trajectories.  These have never arisen before in previous contexts but do arise here precisely because of the tunnel barriers.  Again we can simplify the sum of their contributions using recursion relations and we verify that our final results are sensible with a number of consistency checks.  Finally, we use this to provide an expansion for the conductance in section~\ref{conductance} before presenting our conclusions in section~\ref{conclusions}.

\section{Life under a tunnelling regime} \label{tunnels}

We are considering a chaotic system linked to a lead carrying $M$ scattering channels.  With perfect coupling to the leads, the survival probability of a long (ergodic) stretch, like a link of time $t$, decays exponentially, $\rme^{-\mu t}$, where $\mu=M/T_{\mathrm{H}}$ is the classical escape rate and $T_{\mathrm{H}}$ is the Heisenberg time.  When we consider an encounter which involves $l$ encounter stretches that last $t_{\mathrm{enc}}$ each, because the stretches are close and correlated, if one escapes they all do, so the survival probability depends only on the time of a single crossing of the encounter, $\rme^{-\mu t_{\mathrm{enc}}}$, increasing slightly the survival probability of the trajectory as a whole \cite{heusleretal06}.

When we include tunnel barriers, the situation changes somewhat \cite{whitney07}.  To be precise we add a thin potential wall at the end of the lead so that any incoming (or outgoing) particle is separated into a transmitted and a reflected part.  This simulates imperfect coupling to the leads, a situation which often occurs in real physical systems from quantum dots to microwave billiards, and so makes the theoretical semiclassical treatment of much wider experimental relevance.  In particular we take the limit of tall and narrow barriers, so that hitting the tunnels barriers can be treated as a stochastic event, where a trajectory has probability $p_{m}$ to pass through on hitting channel $m$ (and probability $1-p_{m}$ to be reflected).  The survival probability of a single classical path is then
\beq
\rme^{-\mu_{1} t}, \qquad \mu_{1}=\mu\sum_{m=1}^{M}\frac{p_{m}}{M}.
\eeq
As we consider encounters involving $l$ encounter stretches, we also need to know their joint survival probability.  The probability for all $l$ stretches to survive upon hitting channel $m$ is $(1-p_{m})^l$, so the probability to not survive is simply 1 minus this quantity, leading to an escape rate \cite{whitney07}
\beq \label{muleqn}
\mu_{l}=\mu\sum_{m=1}^{M}\frac{1-(1-p_{m})^l}{M}.
\eeq
With these changes to the survival probability of trajectories, we are now ready to treat the semiclassical approximation to the survival probability of an initially trapped wavepacket.

\section{Survival probability} \label{survprob}

The survival probability of an initially trapped wavepacket is given by
\beq
\rho (t)=\int_A \rmd \r \: \psi(\r,t)\psi^{*}(\r,t),
\eeq
where $A$ is the volume of the corresponding closed system and $\psi(\r,t)$ is the solution of the time dependent Schr\"odinger equation.  As the wavepacket is initially trapped, we have $\rho(0)=1$.  The survival probability is linked to an integrated current density via the continuity equation
\beq
\label{intconteqn}
\frac{\partial }{\partial t}\rho (t)+J(t)=0, \qquad J(t)=\int _{S} \j(\r,t)\cdot \hat{n}_x \rmd x=0,
\eeq
where $S$ is the cross-section of the opening, with $\hat{n}_x $ is the vector normal to this section at the point $x$ in $S$, and
\beq 
\j(\r,t)= \frac{\hbar}{2m\rmi}[\psi^{*}(\r,t)\nabla \psi(\r,t)-\psi(\r,t)\nabla \psi^{*}(\r,t)],
\eeq
is the probability current density.  

The semiclassical approximation to the survival probability $\rho(t)$ was treated in \cite{waltneretal08, gutierrezetal09}, and we use those results as the starting point for our inclusion of tunnel barriers.  The diagonal approximation, which only involves a single trajectory that starts and ends inside the system, leads to the simple result
\beq \label{rhodiageqn}
\rho(t)^{\mathrm{diag}}=\rme^{-\mu_{1} t}.
\eeq

To move beyond the diagonal term we consider trajectories that have close self-encounters, working along the lines of \cite{heusleretal06,mulleretal07}.  As the survival probability was considered in detail in \cite{gutierrezetal09}, we briefly point out the main results here as well as the necessary modifications due to the presence of the tunnel barriers.  Trajectories are labelled by a vector $\v$, whose elements $v_l$ list the number of $l$-encounters along the trajectory, the total number of which is $V=\sum v_{l}$.  The number of links of the related closed periodic orbit is $L=\sum lv_{l}$ from which we can generate the open trajectories by cutting each of those links, meaning that the number of trajectory structures $N(\v)$ corresponding to the vector $\v$ is closely related to the number of closed periodic orbit structures.  The semiclassical contribution is separated into these three cases:
\begin{description}
\item[A] where the start and end points are outside of the encounters,
\item[B] where either the start or the end point is inside an encounter, and 
\item[C] where both the start and end point are inside (different) encounters.
\end{description}

\begin{figure}
\centering
$\begin{array}{ccc}
a) \includegraphics[width=3.5cm]{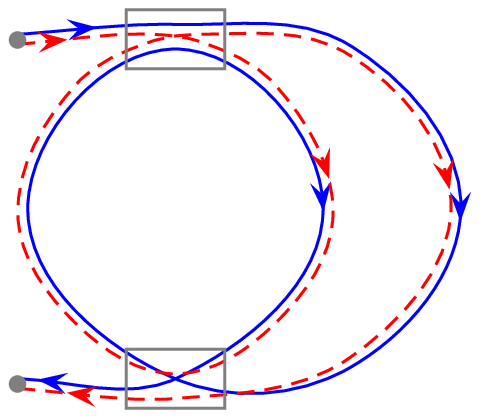}  
& b) \includegraphics[width=3.5cm]{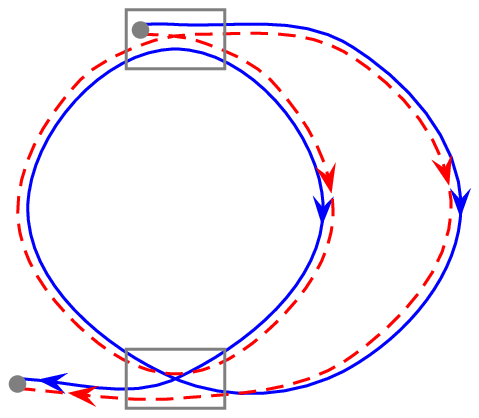}
& c) \includegraphics[width=3.5cm]{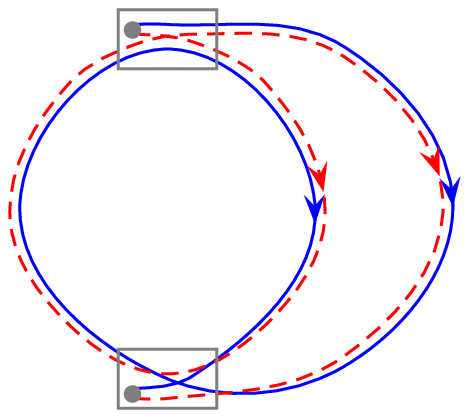} 
\end{array}$
\caption{Examples of the three different types of trajectory structure that contribute to the survival probability.  In (a) both ends are outside of the encounters, in (b) the start point is inside the first encounter while in (c) the end point is also inside an encounter.}
\label{threecasespic}
\end{figure}

These three cases are illustrated in figure~\ref{threecasespic} for a trajectory structure with two 2-encounters.

\subsection{Case A}

To generate the trajectory structures for this case, we simply cut one link of the corresponding closed periodic orbit.  This forms two links in the trajectory structure, giving a total $L+1$ links.  For example, if we cut any of the links of the periodic orbit in figure~\ref{gueencounterpic}a, we arrive at the structure in figure~\ref{threecasespic}a.  From a trajectory $\gamma$, we can create a partner trajectory $\gamma'$ by reconnecting the encounter stretches, leading to an action difference of $\Delta S\approx\s\u$, where the vectors $\s$ and $\u$ contain the appropriate differences, along the stable and unstable manifold respectively, of the original encounter stretches.  The semiclassical contribution of all structures labelled with a common vector $\v$ can be written as
\beq \label{rhoAcontribeqn}
\rho_{\v, \mathrm{A}}(t)=N(\v)\int \rmd\s\:\rmd\u\:w_{\v,\mathrm{A}}(\s,\u,t)\rme^{-\mu t_{\mathrm{exp}}}\rme^{\frac{\rmi}{\hbar}\s\u},
\eeq
where $w_{\v,\mathrm{A}}(\s,\u,t)$ is the weight of such encounters.  The exponential term $\rme^{-\mu t_{\mathrm{exp}}}$ is the average survival probability of the trajectories and requires the corrections described in section~\ref{tunnels}.  We label the $V$ encounters by $\alpha$, which each involve $l_{\alpha}$ encounter stretches that last $t_{\mathrm{enc}}^{\alpha}$.  We further label the $L+1$ links by $i$, which each last $t_{i}$, and then the exposure time is given by
\beq \label{exptimeeqn}
\mu t_{\mathrm{exp}}=\sum_{i=1}^{L+1}\mu_{1}t_{i}+\sum_{\alpha=1}^{V}\mu_{l_{\alpha}}t_{\mathrm{enc}}^{\alpha}= \mu_{1}t - \sum_{\alpha=1}^{V}(\mu_{1}l_{\alpha}-\mu_{l_{\alpha}})t_{\mathrm{enc}}^{\alpha}.
\eeq
To simplify the calculation, we rewrite \eref{rhoAcontribeqn} as
\beq \label{rhoAcontribeqn2}
\rho_{\v, \mathrm{A}}(t)=N(\v)\int \rmd\s\:\rmd\u\:z_{\v,\mathrm{A}}(\s,\u,t)\rme^{-\mu_{1} t}\rme^{\frac{\rmi}{\hbar}\s\u},
\eeq
using an augmented weight, $z_{\v,\mathrm{A}}(\s,\u,t)$, which includes the term from the survival probability of the encounters coming from the right hand side of \eref{exptimeeqn}.
\beqa \label{zeqnA}
z_{\v,\mathrm{A}}(\s,\u,t)&=w_{\v,\mathrm{A}}(\s,\u,t)\rme^{\sum_{\alpha}(\mu_{1}l_{\alpha}-\mu_{l_{\alpha}})t_{\mathrm{enc}}^{\alpha}} \\ \nonumber
&\approx\frac{\left(t-\sum_{\alpha}l_{\alpha}t_{\mathrm{enc}}^{\alpha}\right)^L\prod_{\alpha}\left(1+(\mu_{1}l_{\alpha}-\mu_{l_{\alpha}})t_{\mathrm{enc}}^{\alpha}\right)}{L!\Omega^{L-V}\prod_{\alpha}t_{\mathrm{enc}}^{\alpha}},
\eeqa
where the weight comes from \cite{heusleretal06} and we have expanded the exponent to first order.  From here the semiclassical contribution can be easily found as it only comes from those terms where the encounter times cancel exactly \cite{mulleretal04,mulleretal05}.  The integrals over $\s$ and $\u$ in \eref{rhoAcontribeqn2} provide a factor of $(2\pi\hbar)^{(f-1)(L-V)}$ where $f$ is the number of degrees of freedom of the system.  This factor can be combined with the phase space volume $\Omega$ appearing in \eref{zeqnA} and rewritten in terms of the Heisenberg time as $T_{\mathrm{H}}=\Omega/(2\pi\hbar)^{f-1}$, while the number of structures $N(\v)$ can be found in \cite{muller05b}.

\subsection{Case B}

Starting from the structures for case A, we can also shrink either the link at the start or the end of the trajectory so that it moves into an encounter.  For example, if we shrink the first link of the structure in figure~\ref{threecasespic}a, we arrive at the structure in figure~\ref{threecasespic}b.  This case corresponds to the `one-leg-loops' (1ll) of \cite{waltneretal08}, and we write this contribution as
\beq
\rho_{\v, \mathrm{B}}(t)=N(\v)\int \rmd\s\:\rmd\u\:z_{\v,\mathrm{B}}(\s,\u,t)\rme^{-\mu_{1} t}\rme^{\frac{\rmi}{\hbar}\s\u},
\eeq
where again, once we have the expression for the augmented weight, we can find the semiclassical contribution easily.  Here we have $L$ links in total and an integral over the position of the encounter $\alpha'$ relative to the start or end point.  However, this integral can essentially be replaced by the factor of $t_{\mathrm{enc}}^{\alpha'}$ following a change of variables \cite{gutierrezetal09}.  When we generate the trajectory structures by cutting the links of the corresponding closed periodic orbit we get the encounter $\alpha'$ at the start $l_{\alpha'}$ times, and an equal number of times at the end.  Upon dividing by an overcounting factor $L$, the augmented weight can be simplified to
\beq \label{zeqnB}
\fl z_{\v,\mathrm{B}}(\s,\u,t)\approx\frac{2\left(\sum_{\alpha}l_{\alpha}t_{\mathrm{enc}}^{\alpha}\right)\left(t-\sum_{\alpha}l_{\alpha}t_{\mathrm{enc}}^{\alpha}\right)^{L-1}\prod_{\alpha}\left(1+(\mu_{1}l_{\alpha}-\mu_{l_{\alpha}})t_{\mathrm{enc}}^{\alpha}\right)}{L!\Omega^{L-V}\prod_{\alpha}t_{\mathrm{enc}}^{\alpha}},
\eeq
and we find the contribution as before.

\subsection{Case C}

The final possibility occurs if we additionally shrink the remaining link at the start or the end so that both the start and end point are inside different encounters.  For example, when we shrink the remaining end link of the structure in figure~\ref{threecasespic}b, we arrive at figure~\ref{threecasespic}c, and this is the 0ll case in \cite{gutierrezetal09}.  This contribution can be written as
\beq \label{rhoCcontribeqn}
\rho_{\v, \mathrm{C}}(t)=N(\v)\int \rmd\s\:\rmd\u\:z_{\v,\mathrm{C}}(\s,\u,t)\rme^{-\mu_{1} t}\rme^{\frac{\rmi}{\hbar}\s\u},
\eeq
where we now have $L-1$ links in total and two integrals over the position of the start and end encounters relative to the start and end point.  The number of links of the closed periodic orbit, divided by the factor $L$, which connect encounter $\alpha$ to encounter $\beta$ is denoted $\N_{\alpha,\beta}(\v)$ and recorded in a matrix $\N(\v)$, whose important elements are tabulated in \cite{gutierrezetal09}.  We include the number of possibilities with the augmented weight, which reduces to 
\beqa \label{zeqnC}
\fl N(\v) z_{\v,\mathrm{C}}(\s,\u,t)\approx&\frac{\left(\sum_{\alpha,\beta}\N_{\alpha,\beta}(\v)t_{\mathrm{enc}}^{\alpha}t_{\mathrm{enc}}^{\beta}\right)\left(t-\sum_{\alpha}l_{\alpha}t_{\mathrm{enc}}^{\alpha}\right)^{L-2}}{(L-2)!\Omega^{L-V}\prod_{\alpha}t_{\mathrm{enc}}^{\alpha}}\nonumber\\
&\qquad \times \prod_{\alpha}\left(1+(\mu_{1}l_{\alpha}-\mu_{l_{\alpha}})t_{\mathrm{enc}}^{\alpha}\right),
\eeqa
from which the contribution can be found as before.

\subsection{Unitary case results}

By simply adding the results for the three cases we can obtain the results for each vector and for each symmetry class.  For the unitary case, the first off-diagonal contributions come from a vector with two 2-encounters (which we will denote by $(2)^{2}$), which gives a contribution of
\beq
\rho_{(2)^{2}}(t) = \frac{\rme^{-\mu_{1}t}}{T_{\mathrm{H}}^{2}}\left[\frac{1}{2}t^{2}-\frac{2\mu_{1}-\mu_{2}}{3}t^{3}+\frac{(2\mu_{1}-\mu_{2})^{2}}{24}t^{4}\right],
\eeq
while a vector with a single 3-encounter, denoted $(3)^{1}$, provides
\beq
\rho_{(3)^{1}}(t) = \frac{\rme^{-\mu_{1}t}}{T_{\mathrm{H}}^{2}}\left[-\frac{1}{2}t^{2}+\frac{3\mu_{1}-\mu_{3}}{6}t^{3}\right].
\eeq
We can sum over the different possible trajectory structures, including these ones, and obtain the following expansion for the survival probability
\beqa \label{rhounitaryfulleqn}
\fl \rho(t) = \rme^{-\mu_{1}t}\Bigg[&1-\frac{\mu_{1}-2\mu_{2}+\mu_{3}}{6}\frac{t^{3}}{T_{\mathrm{H}}^{2}}+\frac{(2\mu_{1}-\mu_{2})^{2}}{24}\frac{t^{4}}{T_{\mathrm{H}}^{2}}\nonumber \\
\fl &-\frac{\mu_{1}-4\mu_{2}+2\mu_{3}-4\mu_{4}+\mu_{5}}{15}\frac{t^{5}}{T_{\mathrm{H}}^{4}}+\left(\frac{11\mu_{1}^{2}+14\mu_{2}^{2}+\mu_{3}^{2}-29\mu_{1}\mu_{2}}{60}\right.\nonumber \\
\fl &\qquad\left.+\frac{13\mu_{1}\mu_{3}-3\mu_{1}\mu_{4}}{45}-\frac{7\mu_{2}\mu_{3}}{36}+\frac{\mu_{2}\mu_{4}}{30}\right)\frac{t^{6}}{T_{\mathrm{H}}^{4}}+\dots
\eeqa

However, in order to simplify this general result, we can set all of the individual tunnelling probabilities to $p$.  The Heisenberg time dependence involves only the value of $L-V$ of the vector, and so we further set $t=\tau T_{\mathrm{H}}$.  When we do this for the unitary case we obtain
\beqa
\fl \rho(t) = \rme^{-pM \tau}\Bigg[&1-\frac{p^{2}(p-1)M}{6}\tau^{3}+\frac{p^{4}M^{2}}{24}\tau^{4}-\frac{p^{4}(p-1)M}{15}\tau^{5}+\frac{p^{5}(9p-7)M^{2}}{180}\tau^{6}\nonumber\\
\fl &-\left(\frac{p^{6}(7p-3)M^{3}}{720}+\frac{p^{6}(p-1)M}{28}\right)\tau^{7}\nonumber\\
\fl & +\left(\frac{p^{8}M^{4}}{1920}+\frac{p^{7}(401p-356)M^{2}}{10080}\right)\tau^{8}+\dots
\eeqa
which reduces to the previous result \cite{gutierrezetal09} upon removing the tunnel barrier (by setting $p=1$).  In this form it is easier to see the effect of changing the tunnelling probability, especially if we keep the escape rate (or $pM$) constant.  The first off-diagonal term, which is due to the interplay between trajectories with two 2-encounters and those with a single 3-encounter, actually increases as the tunnelling probability is decreased from 1 (to $1/2$) before falling back to 0 again.  This is due to the fact that the survival probability of a encounter with three correlated stretches falls more slowly than that of an encounter with two stretches leading to an overall increase in the survival probability.  With all the tunnelling probabilities equal, we can compare our semiclassical result with the random matrix result of \cite{ss97} and find full agreement.  The random matrix result is also completely general, with different tunnelling probabilities for each channel, but for the comparison with our semiclassical result we need to perform the integrals in the random matrix result and this only becomes feasible when we set all the tunnelling probabilities equal.  The semiclassical result here in \eref{rhounitaryfulleqn}, of course, provides a direct way of obtaining an expansion for the survival probability also in the case where the tunnel probabilities differ in each channel.  

\subsection{Orthogonal case results}

For the orthogonal case, the first off-diagonal terms comes from trajectories with a single 2-encounter \cite{sr01,rs02}, which give a contribution of
\beq
\rho_{(2)^{1}}(t) = \frac{\rme^{-\mu_{1}t}}{T_{\mathrm{H}}}\frac{2\mu_{1}-\mu_{2}}{2}t^{2}.
\eeq
The next order terms again come from structures with two 2-encounters
\beq
\rho_{(2)^{2}}(t) = \frac{\rme^{-\mu_{1}t}}{T_{\mathrm{H}}^{2}}\left[2t^{2}-\frac{5(2\mu_{1}-\mu_{2})}{3}t^{3}+\frac{5(2\mu_{1}-\mu_{2})^{2}}{24}t^{4}\right],
\eeq
and with a single 3-encounter
\beq
\rho_{(3)^{1}}(t) = \frac{\rme^{-\mu_{1}t}}{T_{\mathrm{H}}^{2}}\left[-2t^{2}+\frac{6\mu_{1}-2\mu_{3}}{3}t^{3}\right].
\eeq
Summing over different structures, we obtain the following expansion
\beqa \label{rhoorthogonalfulleqn}
\fl \rho(t) = \rme^{-\mu_{1}t}\Bigg[&1+\frac{2\mu_{1}-\mu_{2}}{2}\frac{t^{2}}{T_{\mathrm{H}}}-\frac{4\mu_{1}-5\mu_{2}+2\mu_{3}}{3}\frac{t^{3}}{T_{\mathrm{H}}^{2}}+\frac{5(2\mu_{1}-\mu_{2})^{2}}{24}\frac{t^{4}}{T_{\mathrm{H}}^{2}}\nonumber \\
\fl &+\frac{12\mu_{1}-23\mu_{2}+18\mu_{3}-5\mu_{4}}{6}\frac{t^{4}}{T_{\mathrm{H}}^{3}}\nonumber \\
\fl &-\frac{74\mu_{1}^{2}+41\mu_{2}^{2}-119\mu_{1}\mu_{2}+30\mu_{1}\mu_{3}-15\mu_{2}\mu_{3}}{30}\frac{t^{5}}{T_{\mathrm{H}}^{3}}\nonumber \\
\fl &-\frac{96\mu_{1}-250\mu_{2}+297\mu_{3}-168\mu_{4}+37\mu_{5}}{30}\frac{t^{5}}{T_{\mathrm{H}}^{4}}+\dots
\eeqa

We can again set all the tunnelling probabilities to $p$ and simplify the expansion for the orthogonal case, obtaining
\beqa
\fl \rho(t) = \rme^{-pM \tau}\Bigg[&1+\frac{p^{2}M}{2}\tau^{2}-\frac{p^{2}(2p-1)M}{3}\tau^{3}+\left(\frac{5p^{4}M^{2}}{24}+\frac{p^{2}(5p^{2}-2p-1)M}{6}\right)\tau^{4}\nonumber\\
\fl &-\left(\frac{p^{4}(15p-4)M^{2}}{30}+\frac{p^{2}(37p^{3}-17p^{2}-5p-3)M}{30}\right)\tau^{5}\nonumber\\
\fl &+\left(\frac{41p^{6}M^{3}}{720}+\frac{p^{4}(61p^{2}-22p-4)M^{2}}{60}\right.\nonumber\\
\fl &\qquad\left.+\frac{p^{2}(337p^{4}-152p^{3}-51p^{2}-22p-16)M}{180}\right)\tau^{6}\nonumber\\
\fl &-\left(\frac{p^{6}(182p-37)M^{3}}{840}+\frac{p^{4}(2408p^{3}-936p^{2}-231p-65)M^{2}}{1260}\right.\nonumber\\
\fl &\qquad\left.+\frac{p^{2}(1888p^{5}-877p^{4}-266p^{3}-129p^{2}-76p-60)M}{630}\right)\tau^{7}\nonumber\\
\fl &+\dots
\eeqa
which again reduces to the previous result \cite{gutierrezetal09} and agrees with the random matrix result \cite{ss03}, if we transform it following the steps in \cite{ks07b}. Here we can see that reducing the tunnelling probability (at fixed escape rate) leads to a direct reduction of the first off-diagonal term as the enhancement of the survival probability originally due to the closeness of the encounter stretches in the 2-encounter becomes damped.  Without tunnel barriers, the higher order corrections are all due to the slight enhancement to the survival probability that having encounters brings.  But as the tunnel probability is reduced to 0 (at fixed $pM$), this advantage, along with the off-diagonal corrections, vanishes.

Using the same semiclassical techniques, we can also treat transport quantities like the conductance and the reader interested in those results might skip straight to section~\ref{conductance}.  Before we arrive there though, we show in the next three sections how physical properties like continuity arise from semiclassical recursions and how decay is related to transport.  To do this we first move to the Fourier space and consider the survival probability and later its connection to the current density.

\section{Transformed survival probability} \label{fourier}

We start with a trapped wavepacket ($\rho(0)=1$) and restrict ourselves to positive times, so we examine the (one-sided) inverse Fourier transform of the survival probability $\rho(t)$ 
\beq
P(\omega)=\int_{0}^{\infty} \rmd \tau \: \rho(\tau T_{\mathrm{H}}) \rme^{2\pi\rmi\omega\tau},
\eeq
where we still have $\tau=t/T_{\mathrm{H}}$.  As we saw in \cite{kuipersetal08}, the semiclassical contribution for this transformed survival probability can be separated into a simple product of contributions from the encounters and the links, as is possible for the conductance \cite{heusleretal06}.  For example for case A the contribution from structures corresponding to $\v$ is
\beqa
\fl P_{\v,\mathrm{A}}(\omega)=&\frac{N(\v)}{T_{\mathrm{H}}} \left(\prod_{i=1}^{L+1}\int_{0}^{\infty} \rmd t_{i} \: \rme^{-\left(\mu_{1}-\frac{2\pi\rmi\omega}{T_{\mathrm{H}}}\right) t_{i}}\right) \\ \nonumber 
\fl & \qquad \times \left(\prod_{\alpha=1}^{V}\int \rmd\s_{\alpha}\rmd\u_{\alpha}\:\frac{\rme^{-\left(\mu_{l_{\alpha}}-\frac{2\pi\rmi\omega l_{\alpha}}{T_{\mathrm{H}}}\right) t_{\mathrm{enc}}^{\alpha}}\rme^{\frac{\rmi}{\hbar}\s_{\alpha}\u_{\alpha}}}{\Omega^{l_{\alpha}-1}t_{\mathrm{enc}}^{\alpha}}\right),
\eeqa
where we have separated the exposure time using the first expression in \eref{exptimeeqn}.  When we evaluate the integrals, the Heisenberg times cancel meaning that we essentially just obtain a factor of $(G_{1}-2\pi\rmi\omega)^{-1}$ for each link and a factor of $-(G_{l_{\alpha}}-2\pi\rmi\omega l_{\alpha})$ for each encounter, where we define
\beq
G_{l}=\mu_{l}T_{\mathrm{H}}=\sum_{m=1}^{M}\left[1-(1-p_{m})^{l}\right],
\eeq
in line with \eref{muleqn}.  The contribution then becomes
\beq \label{PvAeqn}
P_{\v,\mathrm{A}}(\omega)=N(\v)(-1)^V \frac{\prod_{\alpha}(G_{l_{\alpha}}-2\pi\rmi\omega l_{\alpha})}{(G_{1}-2\pi\rmi\omega)^{L+1}}.
\eeq
Likewise, the diagonal term which involves a single link simply gives
\beq \label{Pdiageqn}
P^{\mathrm{diag}}(\omega)=\frac{1}{(G_{1}-2\pi\rmi\omega)}, 
\eeq
as can be obtained directly from \eref{rhodiageqn}. 

For case B we have one link fewer, leaving $L$ in total, and one encounter, $\alpha'$, at the start or end of the trajectory pair.  This encounter, which occurs $2l_{\alpha'}$ times (divided by $L$), just gives a factor of $1$ leading to the simplified contribution
\beq \label{PvBeqn}
\fl P_{\v,\mathrm{B}}(\omega)=\frac{N(\v)(-1)^{V-1}}{L}\left(\sum_{\alpha'}\frac{2l_{\alpha'}}{(G_{l_{\alpha'}}-2\pi\rmi\omega l_{\alpha'})}\right)\frac{\prod_{\alpha}(G_{l_{\alpha}}-2\pi\rmi\omega l_{\alpha})}{(G_{1}-2\pi\rmi\omega)^{L}}.
\eeq

For case C the two encounters at the end both give factors of $1$ while the remaining encounters and links give their usual contributions, providing
\beq \label{PvCeqn}
\fl P_{\v,\mathrm{C}}(\omega)=\left(\sum_{\alpha,\beta}\frac{\N_{\alpha,\beta}(\v)}{(G_{l_{\alpha}}-2\pi\rmi\omega l_{\alpha})(G_{l_{\beta}}-2\pi\rmi\omega l_{\beta})}\right)\frac{(-1)^{V-2}\prod_{\alpha}(G_{l_{\alpha}}-2\pi\rmi\omega l_{\alpha})}{(G_{1}-2\pi\rmi\omega)^{L-1}}.
\eeq

\subsection{Recursion relations} \label{recursion}

To proceed we consider how we can re-express the sum over vectors with a common value of $L-V$ for case C using recursion relations.  As can be seen in \eref{PvCeqn} it is the sizes of the encounters which is important so we replace $\N_{\alpha,\beta}(\v)$, by $\N_{k,l}(\v)$ which is the number of links connecting a $k$-encounter to an $l$-encounter in all the closed periodic orbits structures described by $\v$ (and divided by $L$).  With this replacement, each vector then gives the following contribution to the survival probability
\beq \label{PvCeqn2}
\fl P_{\v,\mathrm{C}}(\omega)=\left(\sum_{k,l}\frac{\N_{k,l}(\v)}{(G_{k}-2\pi\rmi\omega k)(G_{l}-2\pi\rmi\omega l)}\right)\frac{(-1)^{V-2}\prod_{\alpha}(G_{l_{\alpha}}-2\pi\rmi\omega l_{\alpha})}{(G_{1}-2\pi\rmi\omega)^{L-1}}.
\eeq
Based on the recursion relations in \cite{mulleretal05, muller05b}, the following relation was obtained \cite{kuipersetal08}
\beq \label{twoencorbeqiveqn}
\N_{k,l}(\v)=\frac{k'(v_{k'}+1)}{L-1}N(\v^{[k,l\to k']}),
\eeq
where $k'=k+l-1$, and $v_{k'}$ is the $(k')$-th component of $\v$.  This relation represents merging a $k$- and $l$-encounter in $\v$ to create a $k'$-encounter in $\v^{[k,l\to k']}$ which is correspondingly formed by decreasing the components $v_k$ and $v_l$ by one while increasing the component $v_{k'}$ by one (so that $v_{k'}+1=v_{k'}^{[k,l\to k']}$ for example).  When we include the extra factors to match the form of the survival probability, \eref{twoencorbeqiveqn} becomes
\beqa \label{NGeqiveqn}
&\frac{(-1)^{V}\N_{k,l}(\v)}{(G_{k}-2\pi\rmi\omega k)(G_{l}-2\pi\rmi\omega l)}\frac{\prod_{\alpha}(G_{l_{\alpha}}-2\pi\rmi\omega l_{\alpha})}{(G_{1}-2\pi\rmi\omega)^{L-1}} \nonumber\\
&\qquad = -\frac{k'v_{k'}^{[k,l\to k']}}{(G_{k'}-2\pi\rmi\omega k')}\hat{N}(\v^{[k,l\to k']},G),
\eeqa
where
\beq
\hat{N}(\v,G)=\frac{(-1)^{V}}{L}\frac{\prod_{\alpha}(G_{l_{\alpha}}-2\pi\rmi\omega l_{\alpha})}{(G_{1}-2\pi\rmi\omega)^{L}}N(\v).
\eeq
We now sum \eref{PvCeqn2} over all vectors with the same value of $L-V=m$, and use the recursion relation \eref{NGeqiveqn} to re-express the $m$-th order contribution to the survival probability for case C as
\beq \label{PmCeqn}
P_{m,\mathrm{C}}(\omega)=-\sum_{\v}^{L-V=m}\sum_{k,l}\frac{k'v'_{k'}}{(G_{k'}-2\pi\rmi\omega k')}\hat{N}(\v',G),
\eeq
where $\v'=\v^{[k,l\to k']}$. To form $\v'$ from $\v$ we combined a $k$ and $l$-encounter which reduces both $L$ and $V$ by one, but leaves the value of $L-V=m$ unchanged.  The sum over $\v$ can then effectively be replaced as a sum over $\v'$ itself \cite{mulleretal05, muller05b}.  We later identify this dummy sum variable $\v'$ with $\v$ when we combine this contribution with the contributions from the other cases to obtain \eref{Pmeqn}.  We first recall the contribution of case A from \eref{PvAeqn}
\beq
P_{m,\mathrm{A}}(\omega)=\sum_{\v}^{L-V=m}\frac{L}{(G_{1}-2\pi\rmi\omega)}\hat{N}(\v,G),
\eeq
and from case B from \eref{PvBeqn}, where we re-express the sum over $\alpha$ in terms of the components of the vector $\v$
\beq
P_{m,\mathrm{B}}(\omega)=-\sum_{\v}^{L-V=m}\sum_{l}\frac{2lv_{l}}{(G_{l}-2\pi\rmi\omega l)}\hat{N}(\v,G),
\eeq
so that we can express the total contribution to the survival probability as
\beqa \label{Pmeqn}
\fl P_{m}(\omega)&=&\sum_{\v}^{L-V=m}\left[\frac{L}{(G_{1}-2\pi\rmi\omega)}-\sum_{l}\frac{2lv_{l}}{(G_{l}-2\pi\rmi\omega l)}-\sum_{k,l}\frac{k'v_{k'}}{(G_{k'}-2\pi\rmi\omega k')}\right]\hat{N}(\v,G).\nonumber\\
\fl &&
\eeqa
Moreover, we can simplify the double sum in the third term.  As $k'=k+l-1$ and $k,l\geq 2$ for each value of $k'$ we get $(k'-2)$ copies of that term.  The double sum can then be simplified as 
\beq \label{klsumsimpeqn}
\sum_{k,l}\frac{k'v'_{k'}}{(G_{k'}-2\pi\rmi\omega k')}=\sum_{k'>2}\frac{k'(k'-2)v'_{k'}}{(G_{k'}-2\pi\rmi\omega k')}.
\eeq
We note that the term $k'=2$ in resulting sum is 0, so we can lower the limit of the sum accordingly.  By identifying $k'$ with $l$, we can combine this sum with the sum over $l$ in \eref{Pmeqn}, which becomes
\beq \label{Pmeqnsimp}
P_{m}(\omega)=\sum_{\v}^{L-V=m}\left[\frac{L}{(G_{1}-2\pi\rmi\omega)}-\sum_{l}\frac{l^{2}v_{l}}{(G_{l}-2\pi\rmi\omega l)}\right]\hat{N}(\v,G).
\eeq

This result means that we can effectively replace all of the different types of contributions by this simplified form.  Of course the contribution of individual vectors differs, but for the sum over all vectors with a common value of $L-V=m$, which is the more useful semiclassical quantity, this is a very helpful simplification.

\section{Transformed current density} \label{current}

The current density, which is connected to the survival probability through the continuity equation, is expressed semiclassically in terms of trajectories that start inside the cavity but end in the lead.  Without tunnel barriers, as soon as the trajectory hits the lead it escapes so, as we saw in \cite{kuipersetal08}, we cannot have case C and we only get half the contribution for case B as the encounter can no longer occur at the end.  With tunnel barriers these cases are again possible, as long as the trajectory is reflected on all but the last encounter stretch in the lead.

We consider the Fourier transform of the integrated current density $J(t)$ 
\beq
\J(\omega)=\int_{0}^{\infty} \rmd \tau \: J(\tau T_{\mathrm{H}}) \rme^{2\pi\rmi\omega\tau},
\eeq
for which we can again write the semiclassical contribution as a product of contributions from the links and encounters, as for the survival probability.  Now however, the end of the trajectory must hit the lead and escape.  Whichever channel $m$ it hits, it only has a probability $p_{m}$ of escaping, and this leads to a channel factor of $G_{1}=\sum_{m=1}^{M}p_{m}$.  For the diagonal approximation we then get
\beq \label{Jdiageqn}
\J^{\mathrm{diag}}(\omega)=\frac{1}{T_{\mathrm{H}}} \frac{G_{1}}{(G_{1}-2\pi\rmi\omega)}, 
\eeq
while for trajectory structures described by $\v$ for case A we obtain
\beq \label{JvAeqn}
\J_{\v,\mathrm{A}}(\omega)=\frac{G_{1}}{T_{\mathrm{H}}} P_{\v,\mathrm{A}}(\omega).
\eeq

With case B we have to treat the case where the end of the trajectory is during an encounter differently from the case when the start is during an encounter.  When the encounter occurs at the start, which happens half of the time, we can treat it as we did for the survival probability as long as we include the final escape probability factor $G_{1}$.  When an $l$-encounter occurs at the end it must be reflected the first $(l-1)$ times so as to return to and create the encounter, while escaping on the last encounter stretch.  The probability of this happening for a particular channel is then simply $(1-p_{m})^{l-1}p_{m}$, which we must sum over all channels and include as a factor too.  If we define
\beq
H_{l}=\sum_{m=1}^{M}p_{m}(1-p_{m})^{l-1},
\eeq
then the total contribution can be expressed as
\beq \label{JvBeqn}
\fl \J_{\v,\mathrm{B}}(\omega)=\frac{N(\v)(-1)^{V-1}}{LT_{\mathrm{H}}}\left(\sum_{\alpha'}\frac{l_{\alpha'}\left(G_{1}+H_{l_{\alpha'}}\right)}{(G_{l_{\alpha'}}-2\pi\rmi\omega l_{\alpha'})}\right)\frac{\prod_{\alpha}(G_{l_{\alpha}}-2\pi\rmi\omega l_{\alpha})}{(G_{1}-2\pi\rmi\omega)^{L}}.
\eeq

For case C we know we have an encounter at both ends so, if we say that encounter $\alpha$ is at the lead, while $\beta$ is at the start of the trajectory, we can simplify the result to
\beq \label{JvCeqn}
\fl J_{\v,\mathrm{C}}(\omega)=\left(\sum_{\alpha,\beta}\frac{\N_{\alpha,\beta}(\v)H_{l_{\alpha}}}{(G_{l_{\alpha}}-2\pi\rmi\omega l_{\alpha})(G_{l_{\beta}}-2\pi\rmi\omega l_{\beta})}\right)\frac{(-1)^{V-2}\prod_{\alpha}(G_{l_{\alpha}}-2\pi\rmi\omega l_{\alpha})}{T_{\mathrm{H}}(G_{1}-2\pi\rmi\omega)^{L-1}}.
\eeq
Note that the matrix $\N_{\alpha,\beta}(\v)$ and hence the double sum is symmetric under swapping $\alpha$ and $\beta$.

When we combine the different cases, and re-express the contribution from case C using the recursion relations in section~\ref{recursion}, we obtain
\beqa \label{Jmeqn}
\fl \J_{m}(\omega)&=&\sum_{\v}^{L-V=m}\left[\frac{LG_{1}}{(G_{1}-2\pi\rmi\omega)}-\sum_{l}\frac{lv_{l}(G_{1}+H_{l})}{(G_{l}-2\pi\rmi\omega l)}-\sum_{k,l}\frac{k'v_{k'}H_{k}}{(G_{k'}-2\pi\rmi\omega k')}\right]\frac{\hat{N}(\v,G)}{T_{\mathrm{H}}}.\nonumber\\
\fl&&
\eeqa
Again we can simplify the double sum in the third term.  For each $k'$ we obtain a copy of each $H_{k}$ for $k=2,\dots,k'-1$.  The double sum is then 
\beq \label{klsumsimpeqn2}
\sum_{k,l}\frac{k'v'_{k'}H_{k}}{(G_{k'}-2\pi\rmi\omega k')}=\sum_{k'>2}\frac{k'v'_{k'}\sum_{k=2}^{k'-1}H_{k}}{(G_{k'}-2\pi\rmi\omega k')}.
\eeq
The $k'=2$ term would involve no sum over $H_{k}$ and is formally zero so we can again lower the limit of the sum. We can then combine this sum with the sum over $l$ in \eref{Jmeqn} which becomes
\beq \label{Jmeqnsimptemp}
\fl \J_{m}(\omega)=\sum_{\v}^{L-V=m}\left[\frac{LG_{1}}{(G_{1}-2\pi\rmi\omega)}-\sum_{l}\frac{lv_{l}(G_{1}+\sum_{k=2}^{l-1}H_{k}+H_{l})}{(G_{l}-2\pi\rmi\omega l)}\right]\frac{\hat{N}(\v,G)}{T_{\mathrm{H}}}.
\eeq
We can simplify further because
\beq \label{GHsumsimpeqn}
\fl G_{1}+\sum_{k=2}^{l-1}H_{k}+H_{l}=\sum_{m=1}^{M}p_{m}\sum_{k=1}^{l}(1-p_{m})^{k-1}=\sum_{m=1}^{M}p_{m}\frac{1-(1-p_{m})^{l}}{1-(1-p_{m})}=G_{l},
\eeq
as the sum just involves a geometric progression. The final result for the current density is then
\beq \label{Jmeqnsimp}
 \J_{m}(\omega)=\sum_{\v}^{L-V=m}\left[\frac{LG_{1}}{(G_{1}-2\pi\rmi\omega)}-\sum_{l}\frac{lv_{l}G_{l}}{(G_{l}-2\pi\rmi\omega l)}\right]\frac{\hat{N}(\v,G)}{T_{\mathrm{H}}}.
\eeq

\subsection{Continuity Equation}\label{conteqn}

We have examined the current density because it is connected to the survival probability via the continuity equation \eref{intconteqn}, which in the Fourier space becomes
\beq \label{fourierconteq}
T_{\mathrm{H}}\J(\omega)-(2\pi\rmi\omega)P(\omega)=1,
\eeq
where semiclassically the 1 comes from the diagonal terms, which can easily be checked using \eref{Pdiageqn} and \eref{Jdiageqn}.  As such, to ensure that the continuity equation is satisfied semiclassically, we need to show that the off-diagonal terms vanish and that
\beq \label{contftcompeqn}
T_{\mathrm{H}}\J_{m}(\omega)-(2\pi\rmi\omega) P_{m}(\omega)=0,
\eeq
for all $m>0$.  Combining \eref{Pmeqnsimp} and \eref{Jmeqnsimp}, we have to evaluate
\beq \label{contfteqn}
\fl \sum_{\v}^{L-V=m}\left[\frac{L(G_{1}-2\pi\rmi\omega)}{(G_{1}-2\pi\rmi\omega)}-\sum_{l}\frac{lv_{l}(G_{l}-2\pi\rmi\omega l)}{(G_{l}-2\pi\rmi\omega l)}\right]\hat{N}(\v,G),
\eeq
which directly reduces to
\beq
\sum_{\v}^{L-V=m}\left[L-\sum_{l}lv_{l}\right]\hat{N}(\v,G)=0,
\eeq
since $\sum_{l}lv_{l}=L$.  This verifies \eref{contftcompeqn} and that the semiclassical expansion respects the continuity equation.  Adding the diagonal terms, we indeed obtain \eref{fourierconteq} in our semiclassical regime.

\section{Connection to transport}\label{transport}

A large area of interest in semiclassics is the treatment of quantum transport, rather than decay, but we have followed this route because, as we saw in \cite{kuipersetal08}, the current density is connected to a transport quantity $F(t)$ via the continuity equation $F(t)+\partial J(t)/{\partial t}=0$.  In the Fourier space this continuity equation is
\beq \label{conteqnft2}
T_{\mathrm{H}}\F(\omega)-(2\pi\rmi\omega)\J(\omega)=1,
\eeq
where the semiclassical approximation to $\F(\omega)$ is given by
\beq \label{Ftrajeqn}
\F(\omega) \approx \frac{1}{T_{\mathrm{H}}^{3}}\sum_{a,b} \sum_{\gamma,\gamma'(a\to b)}D_{\gamma}D_{\gamma'}^* \rme^{\frac{\rmi}{\hbar}(S_{\gamma}-S_{\gamma'})}\rme^{\frac{\pi\rmi\omega}{T_{\mathrm{H}}}(t_{\gamma}+t_{\gamma'})},
\eeq
where the sum over $a$ and $b$ is over the channels in the lead and we sum over trajectories $\gamma$ and $\gamma'$ connecting these channels, which have actions $S_{\gamma}$, stability amplitudes $D_{\gamma}$ and times $t_{\gamma}$. This is a semiclassical approximation to a correlation function of scattering matrix elements
\beq 
\F(\omega) \approx \frac{1}{T_{\mathrm{H}}^{2}}\sum_{a,b} S_{ba}\left(E+\frac{\pi\hbar\omega}{T_{\mathrm{H}}}\right)S_{ba}^{*}\left(E-\frac{\pi\hbar\omega}{T_{\mathrm{H}}}\right)
\eeq
which was considered in detail in \cite{ks08} and is related to the Wigner time delay as well as the ac conductance \cite{petitjeanetal08} and hence the conductance. We first calculate the contributions and then we will later provide an expansion for the conductance.

The simplest contribution is the diagonal approximation which reduces to
\beq \label{Fdiageqn}
\F^{\mathrm{diag}}(\omega) \approx \frac{1}{T_{\mathrm{H}}^{3}}\sum_{a,b} \sum_{\gamma(a\to b)}\vert D_{\gamma}\vert^{2} \rme^{\frac{2\pi\rmi\omega}{T_{\mathrm{H}}} t_{\gamma}},
\eeq
where we will use the sum rule \cite{rs02} which has been implicit in our previous calculations
\beq
\sum_{\gamma(a\to b)}\vert D_{\gamma}\vert^{2}\dots \approx \int_{0}^{\infty}\rmd t_{\gamma} \rme^{-\mu_{1}t_{\gamma}}\dots
\eeq
and modified to represent the survival probability now that we have tunnel barriers.  We also need to perform the sum over the channels $a$ and $b$.  We remember that we have a probability of $p_{a}$ of tunnelling into the cavity through channel $a$ and likewise a probability $p_{b}$ of leaving through channel $b$.  The sum over channels is then simply $\sum_{a,b}p_{a}p_{b}=G_{1}^{2}$.  Substituting into \eref{Fdiageqn} and performing the integral we obtain
\beq \label{Fdiageqnend}
\F^{\mathrm{diag}}(\omega) \approx \frac{1}{T_{\mathrm{H}}^{2}}\frac{G_{1}^{2}}{(G_{1}-2\pi\rmi\omega)},
\eeq
where we recall that $G_{1}=\mu_{1}/T_{\mathrm{H}}$.  We can consider that we have an additional contribution for systems with time reversal symmetry when the start and end channels coincide ($a=b$).  Then we can also compare the trajectory $\gamma$ with the time reversal of its partner $\overline{\gamma'}$ giving an additional $p_{a}^{2}$ for this channel combination.  This extra possibility corresponds to coherent backscattering (cbs) and must be considered more carefully when Ehrenfest time effects are important since we actually have a 2-encounter in the lead.  It suits our purposes here to include this contribution with the other contributions involving a 2-encounter, which we explore in section~\ref{caseD}, and not with the diagonal approximation.

We can find the contribution of correlated trajectories using the open sum rule and an auxiliary weight function as before \cite{heusleretal06,mulleretal07} splitting the contribution into encounters and links as for the survival probability in section~\ref{fourier}.  For case A, the contribution of each vector can be written as
\beq \label{FvAconteqn}
\F_{\v,\mathrm{A}}(\omega)= \frac{LG_{1}^{2}}{(G_{1}-2\pi\rmi\omega)}\frac{\hat{N}(\v,G)}{T_{\mathrm{H}}^{2}}.
\eeq
For case B, we have an $l$-encounter at the start or the end so we get a channel factor of $G_{1}H_{l}$, and an additional factor of 2, giving
\beq \label{FvBconteqn}
\F_{\v,\mathrm{B}}(\omega)= -\sum_{l}\frac{2lv_{l}G_{1}H_{l}}{(G_{l}-2\pi\rmi\omega l)}\frac{\hat{N}(\v,G)}{T_{\mathrm{H}}^{2}},
\eeq
while for case C, with an $l$-encounter at the start and a $k$-encounter at the end, we have a channel factor of $H_{k}H_{l}$.  To simplify matters we will also sum over all vectors with the same value of $L-V=m$ and re-express the result using the recursion relations in section~\ref{recursion} to obtain
\beq \label{FvCconteqn}
\F_{m,\mathrm{C}}(\omega)= -\sum_{\v}^{L-V=m}\sum_{k,l}\frac{k'v_{k'}H_{k}H_{l}}{(G_{k'}-2\pi\rmi\omega k')}\frac{\hat{N}(\v,G)}{T_{\mathrm{H}}^{2}},
\eeq
where again $k'=k+l-1$.  As we also saw in section~\ref{recursion}, we can simplify the sum over $k,l$ and combine the result with the sum over $l$ of the contribution of case B, to obtain a total result for cases B and C of
\beq \label{FmBCconteqn}
\F_{m,\mathrm{B}+\mathrm{C}}(\omega)= -\sum_{\v}^{L-V=m}\sum_{l}\frac{lv_{l}\K_{l}}{(G_{l}-2\pi\rmi\omega l)}\frac{\hat{N}(\v,G)}{T_{\mathrm{H}}^{2}},
\eeq
where we have defined
\beq
\K_{l}=\sum_{k=1}^{l}H_{k}H_{l-k+1},
\eeq
and where the $k=1$ and $k=l$ term effectively come from case B (as $G_{1}=H_{1}$), while the rest come from case C. 

However, for transport quantities, where we can start and end in the same channel, there is an additional possibility: Case D, where both the start and end point are inside the same encounter (cbs).

\subsection{Case D} \label{caseD}

When the start and end channel are the same, we have the additional possibility that the trajectory can start and end in the same encounter at the lead.  The coherent backscattering contributions included here which involve a 2-encounter at the lead could be included in the cases above, as for systems with time reversal symmetry we can always add a 2-encounter near the lead without affecting the rest of the trajectory (though for cases B and C this creates a more complicated encounter at the lead).  Although they are included here for practical reason, we are particularly interested in the possibilities that can only occur with tunnel barriers.  These possibilities can also occur for systems without time reversal symmetry, marking their difference from coherent backscattering, though they are in another sense a generalisation of that case.  They are also a 0ll contribution like case C, but they only involve a single encounter at the lead like case B.

To calculate their contribution, we remember that for case C we started with the periodic orbit structures and counted all the links that connected two different encounters.  Obviously, if the link does not connect two different encounters it must connect the encounter to itself, and for all structures of type $\v$ we record these numbers in a vector $\bN(\v)$, where the component $\bN_{l}(\v)$ is the number links connecting an $l$-encounter to itself.  We also need to know the channel factor for such an $l$-encounter in the lead.  We first enter the channel, $a$, with probability $p_{a}$, then get reflected $(l-2)$ times before escaping on the last stretch, giving a total factor of $I_{l-1}$, where
\beq
I_{l}=\sum_{m=1}^{M}p_{m}^{2}(1-p_{m})^{l-1}.
\eeq
There are $L-1$ links in total, and a more complicated integral over the encounter, but the result reduces to
\beq \label{FvDeqn}
\F_{\v,\mathrm{D}}(\omega)=\left(\sum_{l}\frac{\bN_{l}(\v)I_{l-1}}{(G_{l}-2\pi\rmi\omega l)}\right)\frac{(-1)^{V-1}\prod_{\alpha}(G_{l_{\alpha}}-2\pi\rmi\omega l_{\alpha})}{(G_{1}-2\pi\rmi\omega)^{L-1}},
\eeq

Because a link either connects an encounter to itself or a different encounter, we have the relation
\beq \label{bNsumeqn}
N(\v)=\sum_{l}\bN_{l}(\v)+\sum_{k,l}\N_{k,l}(\v).
\eeq
However, we can, using the relations in \cite{mulleretal05,muller05b}, break this sum up into contributions coming from each $l$
\beq \label{bNleqn}
\frac{lv_{l}}{L}N(\v)=\bN_{l}(\v)+\sum_{k}\N_{k,l}(\v),
\eeq
where we note that $\N_{k,l}$ is a symmetric matrix, and if we sum this result over $l$ we recover \eref{bNsumeqn}.  We also note that for systems with time reversal symmetry $\N_{2}(\v)=N(\v^{[2\to]})$ where $\v^{[2\to]}$ is the vector formed by removing a 2-encounter from $\v$.  This relation is the reason why we could also include coherent backscattering involving a 2-encounter at the lead as an additional channel factor in cases A, B and C, but we instead include this contribution here precisely because of \eref{bNleqn} as we already have recursion relations for $\N_{k,l}$.  When we substitute the result \eref{bNleqn} into \eref{FvDeqn}, and use the recursion relation \eref{NGeqiveqn} we obtain a contribution of
\beqa \label{FvDeqn2}
\F_{\v,\mathrm{D}}(\omega)=&-\left[\sum_{l}\frac{lv_{l}I_{l-1}(G_{1}-2\pi\rmi\omega)}{(G_{l}-2\pi\rmi\omega l)}\right]\hat{N}(\v,G)\nonumber\\
&-\left(\sum_{k,l}\frac{k'v'_{k'}I_{l-1}(G_{k}-2\pi\rmi\omega k)}{(G_{k'}-2\pi\rmi\omega k')}\hat{N}(\v',G)\right),
\eeqa
where we recall that $k'=k+l-1$ and $\v'=\v^{[k,l\to k']}$.  When we sum over all vectors with a common value of $L-V=m$ we can re-express the second line, perform one of the sums and combine the result with the first line to obtain
\beq \label{FmDeqn}
\fl \F_{m,\mathrm{D}}(\omega)=-\sum_{\v}^{L-V=m}\left[\sum_{l}\frac{lv_{l}\sum_{k=1}^{l-1}I_{k}\left[G_{l-k}-2\pi\rmi\omega(l- k)\right]}{(G_{l}-2\pi\rmi\omega l)}\right]\hat{N}(\v,G).
\eeq
Using the results which are shown in~\ref{simpchannelsum}, namely
\beqa
\sum_{k=1}^{l-1}(l-k)I_{k}=lG_{1}-G_{l}, \qquad \sum_{k=1}^{l-1}I_{k}G_{l-k}= G_{1}G_{l} - \K_{l},
\eeqa
we can simplify \eref{FmDeqn} to
\beq \label{FmDeqn2}
\fl \F_{m,\mathrm{D}}(\omega)=-\sum_{\v}^{L-V=m}\left[\sum_{l}\frac{lv_{l}\left[G_{1}G_{l}-\K_{l}-2\pi\rmi\omega(lG_{1}- G_{l})\right]}{(G_{l}-2\pi\rmi\omega l)}\right]\hat{N}(\v,G).
\eeq
Finally, we can combine this result with the other cases in \eref{FvAconteqn} and \eref{FmBCconteqn}, to obtain
\beq \label{Fmeqn}
\fl \F_{m}(\omega)=\sum_{\v}^{L-V=m}\left[\frac{LG_{1}^{2}}{(G_{1}-2\pi\rmi\omega)}-\sum_{l}\frac{lv_{l}\left[G_{1}G_{l}-2\pi\rmi\omega(lG_{1}-G_{l})\right]}{(G_{l}-2\pi\rmi\omega l)}\right]\frac{\hat{N}(\v,G)}{T_{\mathrm{H}}^{2}}.
\eeq

\subsection{Consistency} \label{consistency}

As a first check of our results for $\F(\omega)$ in \eref{Fdiageqnend} and \eref{Fmeqn}, we consider the case when $\omega=0$, for which
\beq
\F(0)=\frac{\Tr \left[S(E)S^{\dagger}(E)\right]}{T_{\mathrm{H}}^{2}}.
\eeq
From the diagonal term in \eref{Fdiageqnend}, we obtain
\beq
\F^{\mathrm{diag}}(0)=\frac{G_{1}}{T_{\mathrm{H}}^{2}},
\eeq
while for the off-diagonal terms we have
\beq \label{Fm0eqn}
\F_{m}(0)=\sum_{\v}^{L-V=m}\left[LG_{1}-\sum_{l}lv_{l}G_{1}\right]\frac{(-1)^{V}}{L T_{\mathrm{H}}^{2}}\frac{\prod_{l}G_{l}^{v_{l}}}{G_{1}^{L}},
\eeq
which is identically 0 as $L=\sum_{l}lv_{l}$.  This means that semiclassically we have $\Tr \left[ S S^{\dagger}\right]=G_{1}$, which is the effective number of open channels, and the result we would expect.

Returning to the continuity equation \eref{conteqnft2}, we can now show that
\beqa \label{conteqnmFJ}
&T_{\mathrm{H}}\F_{m}(\omega)-(2\pi\rmi\omega)\J_{m}(\omega)\nonumber\\
& = \sum_{m}\sum_{\v}^{L-V=m}\left[LG_{1}-\sum_{l}\frac{lv_{l}G_{1}(G_{l}-2\pi\rmi\omega l)}{(G_{l}-2\pi\rmi\omega l)}\right]\frac{\hat{N}(\v,G)}{T_{\mathrm{H}}}=0,
\eeqa
for exactly the same reasons as above and in \eref{contfteqn}.  The result in \eref{conteqnmFJ} for all $m>0$ combined with a direct check of the diagonal terms from \eref{Jdiageqn} and \eref{Fdiageqnend}, shows that the continuity equation \eref{conteqnft2} is indeed satisfied in the semiclassical approximation.  

As a final check, we know that $F(\omega)$ is related to the time delay, as in \cite{ks08}, via
\beq
\tau_{\mathrm{W}}=\frac{T_{\mathrm{H}}^{3}}{2\pi\rmi G_{1}}\frac{\rmd}{\rmd \omega}\F(\omega)\bigg\vert_{\omega=0}.
\eeq
If we put in the diagonal approximation result from \eref{Fdiageqnend} into this equation we obtain
\beq \label{wtddiageqn}
\tau_{\mathrm{W}}^{\mathrm{diag}}=\frac{T_{\mathrm{H}}}{G_{1}}=\frac{1}{\mu_{1}},
\eeq
which is exactly the average dwell time of the system with tunnel barriers.  For the off-diagonal terms from \eref{Fmeqn} we obtain
\beqa \label{tauWmeqn}
\fl \tau_{m,\mathrm{W}}=&\frac{1}{\mu_{1}}\sum_{\v}^{L-V=m}\left[L+\sum_{l}\frac{lv_{l}(lG_{1}-G_{l})}{G_{l}}-\sum_{l}l^{2}v_{l}G_{1}\right]\frac{(-1)^{V}}{L}\frac{\prod_{l}G_{l}^{v_{l}}}{G_{1}^{L}}\nonumber\\
\fl & \quad + \frac{1}{\mu_{1}}\sum_{\v}^{L-V=m}\left[LG_{1}-\sum_{l}lv_{l}G_{1}\right]\frac{\rmd}{\rmd \omega}\hat{N}(\v,G)\bigg\vert_{\omega=0} = 0,
\eeqa
again for the same reasons as above.  We therefore find that the average value of our semiclassical expression for the time delay does give the average time delay.  Now that we have shown that the semiclassical results satisfy unitarity and we have a firm footing for the contribution of all the possible cases for chaotic systems with tunnel barriers, we can use these results to calculate a semiclassical expansion for the conductance.

\section{Conductance}\label{conductance}

Up until now we have considered a cavity with a single lead containing $M$ channels.  For the conductance we can imagine splitting this lead into two parts, a left lead containing $M_{\L}$ channels, and a right lead containing $M_{\R}$ channels, where $M_{\L}+M_{\R}=M$.  We will consider an ac conductance given by
\beq 
C(\epsilon) \approx \sum_{a=1}^{M_{\L}}\sum_{b=1}^{M_{\R}} T_{ba}\left(E+\frac{\hbar\epsilon}{2T_{\mathrm{H}}}\right)T_{ba}^{*}\left(E-\frac{\hbar\epsilon}{2T_{\mathrm{H}}}\right),
\eeq
where $T_{ba}$ are the elements of the scattering matrix that connect the left lead to the right lead, and we have set $\epsilon=2\pi\omega$.  The semiclassical approximation is similar to that for $\F(\omega)$, and is given by
\beq \label{Ctrajeqn}
C(\epsilon) \approx \frac{1}{T_{\mathrm{H}}}\sum_{a=1}^{M_{\L}}\sum_{b=1}^{M_{\R}} \sum_{\gamma,\gamma'(a\to b)}D_{\gamma}D_{\gamma'}^* \rme^{\frac{\rmi}{\hbar}(S_{\gamma}-S_{\gamma'})}\rme^{\frac{\rmi\epsilon}{2T_{\mathrm{H}}}(t_{\gamma}+t_{\gamma'})},
\eeq
where the only difference is the change in the channel sum.  The factors for each channel are as before, but the sum is different, so we define
\beq
G_{\L}=\sum_{a=1}^{M_{\L}} p_{a}, \qquad H_{\L,l}=\sum_{a=1}^{M_{\L}} p_{a}(1-p_{a})^{l-1},
\eeq
and similarly for the right lead.  For the diagonal approximation, the channel sum therefore gives a factor $G_{\L}G_{\R}$ and we can write down the result simply as
\beq \label{Cdiageqnend}
C^{\mathrm{diag}}(\epsilon) \approx \frac{G_{\L}G_{\R}}{(G_{1}-\rmi\epsilon)},
\eeq
where we note that because we start and end in different leads there can be no coherent backscattering contribution, and no contribution from case D.  For the rest of case A, the contribution for each vector follows directly as
\beq \label{CvAconteqn}
C_{\v,\mathrm{A}}(\epsilon)= \frac{LG_{\L}G_{\R}}{(G_{1}-\rmi\epsilon)}\hat{N}(\v,G).
\eeq

For case B, if the $l$-encounter occurs at the start, in the left lead, we get a factor $H_{\L,l}G_{\R}$, while if it occurs in the right lead we have the factor $G_{\L}H_{\R,l}$, giving a total contribution of
\beq \label{CvBconteqn}
C_{\v,\mathrm{B}}(\epsilon)= -\sum_{l}\frac{lv_{l}\left[H_{\L,l}G_{\R}+G_{\L}H_{\R,l}\right]}{(G_{l}-\rmi\epsilon l)}\hat{N}(\v,G).
\eeq
For case C, with an $l$-encounter at the start and a $k$-encounter at the end, we have a channel factor of $H_{\L,k}H_{\R,l}$, giving a contribution of
\beq \label{CvCconteqn}
\fl C_{\v,\mathrm{C}}(\epsilon)= \left(\sum_{k,l}\frac{\N_{k,l}(\v)H_{\L,k}H_{\R,l}}{(G_{k}-\rmi\epsilon k)(G_{l}-\rmi\epsilon l)}\right)\frac{(-1)^{V-2}\prod_{\alpha}(G_{l_{\alpha}}-\rmi\epsilon l_{\alpha})}{(G_{1}-\rmi\epsilon)^{L-1}}.
\eeq
With these three cases we can find the semiclassical expansion for the conductance for both symmetry classes.

Further, if we sum over all vectors with a common value of $L-V=m$ for these three cases we can simplify the result to
\beq \label{Cmsimpeqn}
C_{m}(\epsilon)= \sum_{\v}^{L-V=m}\left[\frac{LK_{1}}{(G_{1}-\rmi\epsilon)}-\sum_{l}\frac{lv_{l}K_{l}}{(G_{l}-\rmi\epsilon l)}\right]\hat{N}(\v,G),
\eeq
following the reasoning in section~\ref{transport}, and where for the conductance we define
\beq
K_{l}=\sum_{k=1}^{l}H_{\L,k}H_{\R,l-k+1},
\eeq
and we note that $G_{\L}=H_{\L,1}$ and that $K_{1}=G_{\L}G_{\R}$.  The extremely simple form of the semiclassical result in equation \eref{Cmsimpeqn} is one of the most important results of this paper.  This is the reason why we have re-explored the survival probability, with its three possible cases and its link to the current density via a continuity equation \cite{kuipersetal08}, for the situation where we have tunnel barriers.  For the conductance we have exactly the same three cases, and the recursion relations that we needed to develop for the continuity equation can be applied directly to obtain the simple result \eref{Cmsimpeqn}.  Furthermore, this also simplifies our calculation of the semiclassical expansion of the conductance.

\subsection{Unitary case}

For the unitary case, the first off-diagonal contributions come from a vector with two 2-encounters
\beq
\fl C_{(2)^{2}}(\epsilon) = \frac{G_{\L}G_{\R}(G_{2}-2\rmi\epsilon)^{2}}{(G_{1}-\rmi\epsilon)^{5}} -\frac{(H_{\L,2}G_{\R}+G_{\L}H_{\R,2})(G_{2}-2\rmi\epsilon)}{(G_{1}-\rmi\epsilon)^{4}}+ \frac{H_{\L,2}H_{\R,2}}{(G_{1}-\rmi\epsilon)^{3}},
\eeq
and with a single 3-encounter
\beq
\fl C_{(3)^{1}}(\epsilon) = -\frac{G_{\L}G_{\R}(G_{3}-3\rmi\epsilon)}{(G_{1}-\rmi\epsilon)^{4}} +\frac{(H_{\L,3}G_{\R}+G_{\L}H_{\R,3})}{(G_{1}-\rmi\epsilon)^{3}}.
\eeq
We can then sum over different possible trajectory structures, including these ones, to obtain the expansion
\beqa \label{cunitaryfulleqn}
\fl C(\epsilon) = &\frac{K_{1}}{(G_{1}-\rmi\epsilon)} + \frac{K_{3}}{(G_{1}-\rmi\epsilon)^{3}}-\frac{K_{1}(G_{3}-3\rmi\epsilon)+K_{2}(G_{2}-2\rmi\epsilon)}{(G_{1}-\rmi\epsilon)^{4}}\nonumber \\
\fl & +\frac{K_{1}(G_{2}-2\rmi\epsilon)^{2}+8K_{5}}{(G_{1}-\rmi\epsilon)^{5}}\nonumber \\
\fl & -\frac{8K_{1}(G_{5}-5\rmi\epsilon)+8K_{2}(G_{4}-4\rmi\epsilon)+12K_{3}(G_{3}-3\rmi\epsilon)+16K_{4}(G_{2}-2\rmi\epsilon)}{(G_{1}-\rmi\epsilon)^{6}}\nonumber\\
\fl &+\frac{12K_{1}(G_{3}-3\rmi\epsilon)^{2}+24K_{1}(G_{2}-2\rmi\epsilon)(G_{4}-4\rmi\epsilon)+28K_{2}(G_{2}-2\rmi\epsilon)(G_{3}-3\rmi\epsilon)}{(G_{1}-\rmi\epsilon)^{7}}\nonumber\\
\fl & \quad + \frac{21K_{3}(G_{2}-2\rmi\epsilon)^{2}+180K_{7}}{(G_{1}-\rmi\epsilon)^{7}} + \dots 
\eeqa
This result has not yet been obtained using RMT, but we can compare to previous results if we consider the conductance, which is given by setting $\epsilon=0$.  For the conductance, only the first off-diagonal term is known from RMT \cite{bb96} and this is in agreement with the result here.  To simplify the result further, we set the tunnelling probability in each channel equal to $p$ to obtain
\beq \label{condpunitaryfulleqn}
C(0) = \frac{M_{\L}M_{\R}}{M}\left[p+(p-1)\left(\frac{1}{M^{2}}+\frac{1}{M^{4}}+\frac{1}{M^{6}}+\dots\right)\right],
\eeq
which agrees with the result in \cite{heusleretal06} when we set $p=1$.  Along with the obvious reduction of the conductance as the tunnelling probability is reduced, which reflects the decreased possibility of entering the cavity in the first place, there is also an additional reduction from the higher order terms due to the intricate balance between the survival probabilities of all the different sized encounters.

\subsection{Orthogonal case}

For the orthogonal case, the first off-diagonal contribution comes from a vector with a single 2-encounter
\beq
C_{(2)^{1}}(\epsilon) = -\frac{G_{\L}G_{\R}(G_{2}-2\rmi\epsilon)}{(G_{1}-\rmi\epsilon)^{3}}+\frac{H_{\L,2}G_{\R}+G_{\L}H_{\R,2}}{(G_{1}-\rmi\epsilon)^{2}},
\eeq
while the next order terms are
\beq
\fl C_{(2)^{2}}(\epsilon) = \frac{5G_{\L}G_{\R}(G_{2}-2\rmi\epsilon)^{2}}{(G_{1}-\rmi\epsilon)^{5}} -\frac{5(H_{\L,2}G_{\R}+G_{\L}H_{\R,2})(G_{2}-2\rmi\epsilon)}{(G_{1}-\rmi\epsilon)^{4}}+ \frac{4H_{\L,2}H_{\R,2}}{(G_{1}-\rmi\epsilon)^{3}},
\eeq
and
\beq
\fl C_{(3)^{1}}(\epsilon) = -\frac{4G_{\L}G_{\R}(G_{3}-3\rmi\epsilon)}{(G_{1}-\rmi\epsilon)^{4}} +\frac{4(H_{\L,3}G_{\R}+G_{\L}H_{\R,3})}{(G_{1}-\rmi\epsilon)^{3}}.
\eeq
When we then sum over different possible trajectory structures, we obtain the following expansion
\beqa \label{corthogonalfulleqn}
\fl C(\epsilon) = &\frac{K_{1}}{(G_{1}-\rmi\epsilon)} +\frac{K_{2}}{(G_{1}-\rmi\epsilon)^{2}}-\frac{K_{1}(G_{2}-2\rmi\epsilon)-4K_{3}}{(G_{1}-\rmi\epsilon)^{3}}\nonumber \\
\fl & -\frac{4K_{1}(G_{3}-3\rmi\epsilon)+5K_{2}(G_{2}-2\rmi\epsilon)-20K_{4}}{(G_{1}-\rmi\epsilon)^{4}}\nonumber \\
\fl & +\frac{5K_{1}(G_{2}-2\rmi\epsilon)^{2}-20K_{1}(G_{4}-4\rmi\epsilon)-24K_{2}(G_{3}-3\rmi\epsilon)}{(G_{1}-\rmi\epsilon)^{5}}\nonumber \\
\fl& \quad -\frac{36K_{3}(G_{2}-2\rmi\epsilon)-148K_{5}}{(G_{1}-\rmi\epsilon)^{5}}\nonumber \\
\fl& +\frac{60K_{1}(G_{2}-2\rmi\epsilon)(G_{3}-3\rmi\epsilon)-148K_{1}(G_{5}-5\rmi\epsilon)+41K_{2}(G_{2}-2\rmi\epsilon)^{2}}{(G_{1}-\rmi\epsilon)^{6}}\nonumber\\
\fl&\quad-\frac{168K_{2}(G_{4}-4\rmi\epsilon)+228K_{3}(G_{3}-3\rmi\epsilon)+336K_{4}(G_{2}-2\rmi\epsilon)+1348K_{6}}{(G_{1}-\rmi\epsilon)^{6}}\nonumber\\
\fl &+\dots 
\eeqa
This result is also novel and the conductance can again be obtained by setting $\epsilon=0$.  For the conductance, we can compare the first off-diagonal term with the random matrix result from \cite{bb96}, as well as with the semiclassical result of \cite{whitney07}, and find agreement.  We can look at the simplified case where the tunnelling probability in each channel is equal to $p$ and obtain
\beqa \label{condporthogonalfulleqn}
\fl C(0) = \frac{M_{\L}M_{\R}}{M}&\left[p-\frac{p}{M}+\frac{3p-2}{M^{2}}-\frac{5p^{2}-4}{pM^{3}}+\frac{11p^{3}-6p^{2}+8p-12}{p^{2}M^{4}}\right.\nonumber\\
\fl & \quad -\frac{21p^{4}-4p^{3}-36p^{2}+84p-64}{p^{3}M^{5}}\nonumber\\
\fl & \quad \left.+\frac{43p^{5}-18p^{4}+88p^{3}-528p^{2}+896p-480}{p^{4}M^{6}}+\dots \right]
\eeqa
which also reduces to the result in \cite{heusleretal06} when we remove the tunnel barriers by setting $p=1$.  Considering the conductance as the tunnelling probability is reduced (at fixed $pM$), we can see that the dip in the conductance from the first off-diagonal term also reduces.  This dip compensates the enhancement to the reflectance due to coherent backscattering and, as noted in \cite{whitney07}, as the tunnelling probability decreases these trajectories trying to return to their starting lead are increasingly likely to be reflected back into the system and contribute to the conductance instead.  At higher orders the mechanisms are more complicated, and with new coherent backscattering possibilities (case D) can lead to changes in sign, but the overall reduction of the terms (as $p\to 0$) similarly represents that as trajectories become less likely to leave each time they hit a channel, they become more likely to leave through either lead.

It is worth remarking here that in the small $p$ limit, in particular where $pM\lesssim 1$ the average trajectory spends longer in the cavity than the Heisenberg time, as can be seen from \eref{wtddiageqn}.  The semiclassical treatment presented here is then no longer complete in that regime, but by the same token, adding tunnel barriers leads to an interesting example of where Heisenberg time effects become important in open systems.  Such systems could therefore be useful for exploring classical trajectory correlations beyond the Heisenberg time.

\section{Conclusions} \label{conclusions}

The main focus of this paper has been on the combinatorial relations that connect the different possible trajectory structures and their contributions.  These build on the connections in \cite{kuipersetal08}, where without tunnel barriers we move in a strict hierarchy from a scattering matrix correlation function involving only case A, through the current density where we add case B, and finally to the survival probability where all three cases are allowed (A, B and C).  Upon the addition of tunnel barriers, this hierarchy is equalised and all three cases contribute for all quantities.  Moreover, when we start and end in the same channel, a fourth and new type of contribution appears: type D which is a generalisation of coherent backscattering.  This type of structure was previously ruled out, but from \eref{bNsumeqn} for example we can see how it naturally completes the set.

Along with the combinatorics for the trajectory structures, which can all be generated from the related closed periodic orbit structures by cutting appropriate links, the tunnel barriers add probabilistic combinatorics from the channel factors and escape probabilities \cite{whitney07}.  We have shown how these all combine and simplify mathematically, in non-obvious ways, so that through the semiclassical approximation they give results that respect required physical properties such as continuity.  Simple physics is mirrored by the complicated combinatorial structure, here made more complicated by the tunnel barriers.  The semiclassical methods \cite{heusleretal06,mulleretal07} we used modified appropriately, not only allow us to perform a semiclassical expansion of quantum quantities like the survival probability, as RMT also allows, but give an intuitive explanation in terms of classical trajectories.

In this paper we presented an (almost) complete picture of the semiclassical treatment of chaotic systems with tunnel barriers, for arbitrarily complicated classically correlated trajectories and for a variety of different quantities.  However, the examples covered only involved pairs of trajectories, and it is worth remembering that when we treat quantities that involve more trajectories additional phases can occur for trajectories that never enter the system, as described in \cite{whitney07}.  Moreover, this picture is only complete, and all the results in this paper are only valid, in the particular regime where Ehrenfest time effects are negligible and for times shorter than the Heisenberg time.  But by placing the possible trajectory structures, their contributions and their relations to each other on a firm footing, we hope this work could be a stepping stone to a complete expansion based semiclassical treatment of the Ehrenfest time regime.  Of course the lower order terms have been successfully treated semiclassically \cite{jw06,rb06,wj06,br06}, including tunnel barriers \cite{whitney07}, and these works show that additional trajectory structures appear and contribute in this regime.  These additional structures and the higher order terms remain to be generalised, but hopefully there is some mathematical structure behind these contributions that again can be simplified, like in this article, to reproduce continuity and other physical properties.  Moving to times beyond the Heisenberg time is possibly a much bigger challenge, but progress on this front has already been made \cite{heusleretal07,mulleretal09,km07}.

\ack{The author would like to thank Daniel Waltner, Cyril Petitjean and Klaus Richter for helpful discussions and gratefully acknowledges the Alexander von Humboldt Foundation for funding.}

\appendix

\section{Tables of $\bN_{l}(\v)$}\label{Nlvtables}

Here we record the numbers $\bN_{l}(\v)$ which are useful for calculating semiclassical expansions for transport quantities, like the reflectance, which involve case D.  For the unitary case, a 2-encounter cannot return directly to itself, so $\bN_{2}(\v)=0$, while the remaining numbers are in table~\ref{guenumbers}.  The second column was in \cite{mulleretal05,muller05b}, and is only included for reference, while we note that the numbers needed for case C were included in \cite{gutierrezetal09}.
\Table{\label{guenumbers}The number of trajectory pairs and the number of links connecting encounters to themselves for systems without time reversal symmetry.}
\begin{tabular}{ccccccc}
\br
$\v$&$N(\v)$&$\bN_{3}(\v)$&$\bN_{4}(\v)$&$\bN_{5}(\v)$&$\bN_{6}(\v)$&$\bN_{7}(\v)$\\
\mr
$(2)^{2}$&1&&&&&\\
$(3)^{1}$&1&1&&&&\\
\mr
$(2)^{4}$&21&&&&&\\
$(2)^{2}(3)^{1}$&49&5&&&&\\
$(2)^{1}(4)^{1}$&24&&8&&&\\
$(3)^{2}$&12&4&&&&\\
$(5)^{1}$&8&&&8&&\\
\mr
$(2)^{6}$&1485&&&&&\\
$(2)^{4}(3)^{1}$&5445&189&&&&\\
$(2)^{3}(4)^{1}$&3240&&336&&&\\
$(2)^{2}(3)^{2}$&4440&392&&&&\\
$(2)^{2}(5)^{1}$&1728&&&420&&\\
$(2)^{1}(3)^{1}(4)^{1}$&2952&168&392&&&\\
$(3)^{3}$&464&84&&&&\\
$(2)^{1}(6)^{1}$&720&&&&360&\\
$(3)^{1}(5)^{1}$&608&48&&200&&\\
$(4)^{2}$&276&&96&&&\\
$(7)^{1}$&180&&&&&180\\
\br
\end{tabular}
\endTable

For the orthogonal case, we can always add a 2-encounter at the start and end of the trajectory so $\bN_{2}(\v)=N(\v^{[2\to]})$, where $\v^{[2\to]}$ is the vector formed from $\v$ by removing a 2-encounter.  This can be seen, along with the remaining numbers in table~\ref{goenumbers}, where the second column is from \cite{mulleretal05,muller05b,gutierrezetal09}.  The numbers needed for case C were included in \cite{gutierrezetal09}. 
\Table{\label{goenumbers}The number of trajectory pairs and the number of links connecting encounters to themselves for systems with time reversal symmetry.}
\hspace{-0.6cm}
\begin{tabular}{cccccccc}
\br
$\v$&$N(\v)$&$\bN_{2}(\v)$&$\bN_{3}(\v)$&$\bN_{4}(\v)$&$\bN_{5}(\v)$&$\bN_{6}(\v)$&$\bN_{7}(\v)$\\
\mr
$(2)^{1}$&1&1&&&&&\\
\mr
$(2)^{2}$&5&1&&&&&\\
$(3)^{1}$&4&&4&&&&\\
\mr
$(2)^{3}$&41&5&&&&&\\
$(2)^{1}(3)^{1}$&60&4&16&&&&\\
$(4)^{1}$&20&&&20&&&\\
\mr
$(2)^{4}$&509&41&&&&&\\
$(2)^{2}(3)^{1}$&1092&60&132&&&&\\
$(2)^{1}(4)^{1}$&504&20&&188&&&\\
$(3)^{2}$&228&&80&&&&\\
$(5)^{1}$&148&&&&148&&\\
\mr
$(2)^{5}$&8229&509&&&&&\\
$(2)^{3}(3)^{1}$&23160&1092&1592&&&&\\
$(2)^{2}(4)^{1}$&12256&504&&2388&&&\\
$(2)^{1}(3)^{2}$&10960&228&1968&&&&\\
$(2)^{1}(5)^{1}$&5236&148&&&2392&&\\
$(3)^{1}(4)^{1}$&4396&&536&1164&&&\\
$(6)^{1}$&1348&&&&&1348&\\
\mr
$(2)^{6}$&166377&8229&&&&&\\
$(2)^{4}(3)^{1}$&579876&23160&25620&&&&\\
$(2)^{3}(4)^{1}$&331320&12256&&39448&&&\\
$(2)^{2}(3)^{2}$&443400&10960&48352&&&&\\
$(2)^{2}(5)^{1}$&167544&5236&&&43724&&\\
$(2)^{1}(3)^{1}(4)^{1}$&280368&4396&19312&42132&&&\\
$(3)^{3}$&41792&&8672&&&&\\
$(2)^{1}(6)^{1}$&65808&1348&&&&34252&\\
$(3)^{1}(5)^{1}$&52992&&4768&&18016&&\\
$(4)^{2}$&24788&&&9684&&&\\
$(7)^{1}$&15104&&&&&&15104\\
\br
\end{tabular}
\endTable

\section{Simplifying channel sum products}\label{simpchannelsum}

The first sum we wish to simplify is
\beq
Y=\sum_{k=1}^{l-1}(l-k)I_{k}=\sum_{k=1}^{l-1}\sum_{m}(l-k)p_{m}^{2}(1-p_{m})^{k-1},
\eeq
which just involves geometric progressions.  The first part is simply
\beq
l\sum_{m}p_{m}^{2}\sum_{k=1}^{l-1}(1-p_{m})^{k-1}=l\sum_{m}p_{m}\left[1-(1-p_m)^{l-1}\right],
\eeq
while the second is
\beq
\fl -\sum_{m}p_{m}^{2}\sum_{k=1}^{l-1}k(1-p_{m})^{k-1}=-\sum_{m}\left[1-(1-p_m)^{l}\right]+l\sum_{m}p_{m}(1-p_m)^{l-1}.
\eeq
Combining the two parts, the original sum reduces to
\beqa
Y=l\sum_{m}p_{m}-\sum_{m}\left[1-(1-p_m)^{l}\right]\nonumber\\
=lG_{1}-G_{l}=\sum_{k=1}^{l-1}(l-k)I_{k}.
\eeqa

We also wish to simplify the following sum
\beq \label{Zeqn1}
Z=\K_{l}+\sum_{k=1}^{l-1}I_{k}G_{l-k}=\sum_{k=1}^{l}H_{k}H_{l-k+1}+\sum_{k=1}^{l-1}I_{k}G_{l-k}.
\eeq
The first step is to notice that
\beqa
\fl H_{k+1}&=\sum_{m}p_{m}(1-p_{m})^{k}=\sum_{m}p_{m}(1-p_{m})^{k-1}-\sum_{m}p_{m}^{2}(1-p_{m})^{k-1}\nonumber\\
\fl &=H_{k}-I_{k},
\eeqa
so that we can replace all the $I_{k}$ terms in \eref{Zeqn1} and obtain
\beqa \label{Zeqn2}
Z&=\sum_{k=1}^{l}H_{k}H_{l-k+1}+\sum_{k=1}^{l-1}H_{k}G_{l-k}-\sum_{k=1}^{l-1}H_{k+1}G_{l-k}\nonumber\\
&=H_{1}H_{l}+\sum_{k=1}^{l-1}H_{k}\left(H_{l-k+1}+G_{l-k}\right)-\sum_{k=1}^{l-1}H_{k+1}G_{l-k}.
\eeqa
The next step is the simplification,
\beqa
\fl H_{l-k+1}+G_{l-k}&=\sum_{m}\left(p_{m}(1-p_{m})^{l-k}+1-(1-p_{m})^{k-1}\right)=\sum_{m}\left(1-(1-p_{m})^{l-k}\right)\nonumber\\
\fl &=G_{l-k+1},
\eeqa
which we can put into \eref{Zeqn2}.  We also change the sum index on the second sum (with $k'=k+1$), to obtain  
\beqa \label{Zeqn3}
Z&=H_{1}H_{l}+\sum_{k=1}^{l-1}H_{k}G_{l-k+1}-\sum_{k'=2}^{l}H_{k'}G_{l-k'+1}\nonumber\\
&=H_{1}H_{l}+H_{1}G_{l}-H_{l}G_{1},
\eeqa
since all the terms in the two sums mutually cancel apart from the start and end ones.  As $H_{1}=G_{1}$, we obtain the final result of
\beq \label{Zeqnfinal}
Z=G_{1}G_{l}=\K_{l}+\sum_{k=1}^{l-1}I_{k}G_{l-k}.
\eeq

\end{document}